\documentclass[preprint, aps,prd,floatfix]{revtex4-2}
\usepackage{graphicx}
\usepackage{amsfonts}
\usepackage{amsmath}
\usepackage{rotating}
\usepackage{amssymb,amsthm}

\newtheorem{thm}{Theorem}[section]

\theoremstyle{definition} 
\newtheorem{defn}[thm]{Definition}  
\theoremstyle{remark}  
\newtheorem{rem}[thm]{Remark}  

\newcommand{\mbold}[1]{\mbox{\boldmath{\ensuremath{#1}}}}

\def\beq{\begin{eqnarray}}  
\def\eeq{\end{eqnarray}}  

\def \a {\alpha}
\def \b {\beta}
\def \e {\gamma}
\def \bell {\mbox{{\mbold\ell}}}
\def \bn {\mbox{{\bf n}}}
\def \bm {\mbox{{\bf m}}}
\def \bh {\mbox{{\bf h}}}
\def \bk {\mbox{{\bf k}}}
\def \tv {\tilde{ V}}
\def \bomega {\mbox{{\mbold \omega}}}

\begin{document}

%%-------------------------------------------------------------------------
%%----------------------   Top Matter    ----------------------------------
%%-------------------------------------------------------------------------

\title{Locally-homogeneous Riemann-Cartan geometries with the largest symmetry group}

\author {D. D. McNutt}
\email{mcnuttdd@gmail.com}
\affiliation{ Center for Theoretical Physics, Polish Academy of Sciences, Warsaw, Poland}

\author{R. J. \surname{van den Hoogen}}
\email{rvandenh@stfx.ca}
\affiliation{Department of Mathematics and Statistics, St. Francis Xavier University, Antigonish, Nova Scotia, Canada, B2G 2W5}

\author {A. A. Coley}
\email{alan.coley@dal.ca }
\affiliation{Department of Mathematics and Statistics, Dalhousie University, Halifax, Nova Scotia, Canada, B3H 3J5}

\date{\today}

%%-------------------------------------------------------------------------
%%----------------------   Abstract      ----------------------------------
%%-------------------------------------------------------------------------

\begin{abstract}

The symmetry frame formalism is an effective tool for computing the symmetries of a Riemann-Cartan geometry and, in particular, in metric teleparallel geometries. In the case of non-vanishing torsion in a four dimensional Riemann-Cartan geometry, the Minkowski geometry is the only geometry admitting ten affine frame symmetries.  Excluding this geometry, the maximal number of affine frame symmetries is seven. A natural question is to ask what four dimensional geometries admit a seven-dimensional group of affine frame symmetries. Such geometries are locally homogeneous and admit the largest isotropy group permitted, and hence are called maximally isotropic. Using the symmetry frame formalism to compute affine frame symmetries along with the additional structure of the torsion tensor, we employ the Cartan-Karlhede algorithm to determine all possible seven-dimensional symmetry groups for Riemann-Cartan geometries.

\end{abstract}

\maketitle

%\textsf{•}\tableofcontents 
%%-------------------------------------------------------------------------
%%----------------------   SECTION       ----------------------------------
%%-------------------------------------------------------------------------
\section{Introduction} \label{sec:intro}

Symmetries and symmetry methods are extremely helpful in the study of complex mathematical and physical problems. Mathematicians Lie and Cartan, amongst others, provided a rigourous mathematical framework in which to study symmetries in a geometrical setting. It was Noether who discovered the paramount property linking continuous symmetries of physical systems and conservation laws or constants of motion. With the successful geometrization of gravity via Einstein's General Relativity (GR) these symmetry methods were then catapulted to a more privileged position in theoretical physics. When Glashow, Salam and Weinberg developed their theory of the electroweak interactions \cite{bilenky1982glashow}, the usefulness and importance of symmetry and symmetry methods in the development of fundamental physics was made abundantly clear.

Symmetry techniques have been used successfully in GR to construct tractable systems of governing differential equations suitable for analysis and the determination of solutions.  Some of the first solutions in GR involved the assumption of a static and spherically symmetric geometry.  This geometrical symmetry places constraints on the metric, the geometrical object that encodes gravity in GR. The resulting geometrical framework was then put into the vacuum field equations of GR resulting in the well known Schwarschild black-hole solution.

Any continuous symmetry can be generated by an infinitesimal vector field.  If one assumes the metric is invariant under a continuous symmetry, the Lie derivative of the metric along the vector generator is zero.  These  so-called Killing symmetries are well understood in the literature and have been very successful in generating a plethora of solutions in  GR \cite{kramer}.

 GR is formulated within a geometrical framework in which certain geometrical objects are assumed to be zero; in particular, the torsion and non-metricity are identically zero.  The only remaining non-trivial characteristic of the geometry is the curvature, which can be computed directly from the metric via the Levi-Civita connection, thereby elevating the metric as the fundamental object of study in  GR.  Gravitational models based on  GR have been constructed and experimentally verified. For example, all solar system tests of gravity agree experimentally with  GR \cite{Will:2014kxa}. The observations of the Hulse-Taylor binary pulsar (and others) \cite{Weisberg:2016jye} and the more recent observations of gravitational waves \cite{LIGOScientific:2016aoc,LIGOScientific:2018mvr} aligns extremely well with  GR in the strong gravity regime.  At astrophysical scales, GR satisfies all experimental tests to date.

Unfortunately, as one moves to cosmological scales, one finds that  GR without modification does not adequately explain all cosmological observations.  For example, cosmological models constructed using GR cannot adequately explain the currently observed acceleration of the universe without modification.  The $\Lambda$CDM model within the GR framework fits the cosmological data extremely well \cite{Planck:2018vyg,WMAP:2012nax}, but it does require that we make a subtle but important modification of GR to achieve this level of experimental agreement with the addition of Dark Energy. Within the last few years, more problematic experimental discrepancies have come to light \cite{Buchert}. Observations of the current value of the Hubble expansion parameter using two different observational techniques and, more importantly, different length scales produce two disparate values of the Hubble parameter that can no longer be explained away by systematical error \cite{Riess:2016jrr, Ananthaswamy:2018oh, DiValentino:2022fjm, Hu:2023jqc,coleyellis} and which is made worse by JWST observations \cite{ColeyJWST2023}. This Hubble tension coupled with the already modified GR needed to describe the acceleration of the universe, suggests that maybe one should investigate alternatives to GR.  Perhaps there is an alternative to GR that explains the cosmological disparities while also reproducing GR in some limiting sense.

GR is formulated in the language Riemannian geometry in which symmetry techniques are standard tools. The real challenge in studying an alternative theory of gravity is that symmetry tools and techniques that have been so successful in investigating GR have not been fully developed or generalized. The idea of symmetries and more importantly the application of symmetries in these more general geometries is under current development \cite{Hohmann:2019nat, Coley:2019zld, Hohmann:2021ast, Pfeifer:2022txm, McNutt:2023nxmnxm, obukhov2023poincare}. Our focus in this paper is to apply symmetry techniques to establish the strongest restrictions on the geometrical framework in Rieman-Cartan geometries. In a Riemann-Cartan geometry $(M,{\bf g}, \bomega)$, the differentiable manifold, $M$, is endowed with a connection, $\bomega$, that is compatible with the metric ${\bf g}$, $\nabla {\bf g} = 0 $, but the torsion tensor and curvature tensor are non-trivial.

In particular, it has already been shown that the only Riemann-Cartan geometry that admits a maximal number of symmetries is Minkowski space \cite{Coley:2019zld}.  Furthermore, with the exception of Minkowski space where the torsion tensor and curvature tensor vanish a non-trivial Riemann-Cartan geometry admits at most a seven-dimensional affine frame symmetry group \cite{Coley:2019zld}. These geometries admit the largest possible isotropy subgroup for Riemann-Cartan geometries, and hence will be called {\it maximally isotropic}. In this paper we determine the necessary form of the geometrical inputs, so that the frame (or co-frame), and connection satisfy the seven-dimensional affine frame symmetry constraints.  The resulting geometrical inputs will provide suitable ansatzes for the further investigation of gravitational and cosmological models in wide classes of Riemann-Cartan gravity theories. 

The layout of the paper is as follows. In the remainder of section \ref{sec:intro} we summarize the notation used in the paper. In section \ref{sec:BasicRCG}, we briefly review Riemann-Cartan geometries and discuss the decomposition of the curvature tensor and the torsion tensor. In section \ref{sec:CKalg}, we summarize the Cartan-Karlhede algorithm and derive necessary conditions for a locally homogeneous maximally isotropic geometry. In section \ref{sec:FSs}, we introduce the symmetry frame formalism. In section \ref{sec:3isospaces}, we employ the symmetry frame formalism to determine all spaces that admit $SO(3)$, $SO(1,2)$ and $E(2)$ isotropy subgroups. In section \ref{sec:G4spaces}, we compute the most general $G_4$ groups that admit $SO(3)$, $SO(1,2)$ and $E(2)$ as isotropy groups. In section \ref{sec:G7analysis}, we analyze the conditions introduced in section \ref{sec:CKalg} and determine all locally homogeneous maximally isotropic geometries. The results of this analysis are summarized in section \ref{sec:summary}, and the implications of these results are discussed in \ref{sec:Discussion}.

\subsection{Notation}

We will denote the coordinate indices by $\mu, \nu, \ldots$ and the tangent space indices by $a,b,\ldots$. Unless otherwise indicated the spacetime coordinates will be $x^\mu$. The frame basis elements will be denoted as $\bh_a$ while the dual coframe elements are $\bh^a$. In terms of the coordinate basis, the vierbein components are $h_a^{~\mu}$ or $h^a_{~\mu}$. The metric will be denoted as $g_{\mu \nu}$ in the coordinate basis. Relative to the coframe basis the metric will be the Minkowski tangent space metric which is denoted as $\eta_{ab}$. 

A local Lorentz transformation which leaves $\eta_{ab}$ unchanged, will be written as $\Lambda_a^{~b}(x^\mu)$. The inverse transformation of $\Lambda_a^{~b}(x^\mu)$ is then $\Lambda^a_{~b} = (\Lambda^{-1})_a^{~b}$. The connection one-form, $\bomega^a_{~b}$, is designated by $\bomega^a_{~b} = \omega^a_{~bc} \bh^c$. We will denote the curvature and torsion tensors, respectively, by  $R^a_{~bcd}$ and $T^a_{~bc}$.

In the frame basis, $\{ \bh_a \}$, covariant derivatives with respect to frame members will be written as $\nabla_a \equiv \nabla_{\bh_a}$. As a helpful short-hand, covariant derivatives with respect to a metric-compatible connection will also be written using a semi-colon; i.e.,  $ \nabla_e T_{abc} = T_{abc;e}$. When considering the integral curves of a vector field ${\bf X}$, with parameter $\tau$, given as $\phi_{\tau} : M \to M$, we will write the corresponding Lorentz frame transformation as $\phi_\tau^{*} \bh^a = \Lambda^a_{(\tau)~b} \bh^b$. 

We will use different indices to distinguish the commutator relations of the various quantities used in the symmetry frame formalism. For the affine frame symmetry generators, $\{{\bf X}_I \}$, we will use $[{\bf X}_I, {\bf X}_J] = C^K_{~IJ} {\bf X}_K$. The representation of the isotropy group Lie algebra, $ \lambda_{\hat{i}} =\lambda^a_{~\hat{i} b}$ have the commutator relations, $[\lambda_{\hat{i}}, \lambda_{\hat{j}}] = C^{\hat{k}}_{~\hat{i} \hat{j}} \lambda_{\hat{k}}$. Lastly, the frame basis commutator relations are $[\bh_a, \bh_b] = c^c_{~ab} \bh_c$. 

\section{Riemann-Cartan geometries}\label{sec:BasicRCG}

Let $M$ be a four-dimensional (4D) manifold with coordinates $x^\mu$. The differentiability structure of $M$ ensures there is a non-degenerate coframe field $\bh^a$ defined on a subset $U \subset M$. To determine angles and lengths, and hence define the geometry, we assume $M$ is equipped with a symmetric metric field, $g_{ab}$ derived from the coframe field basis. Lastly, in order to define covariant differentiation of tensor fields, a notion of parallel transport is necessary and the existence of a linear affine connection one-form $\bomega^a_{~b}$ is assumed. 

Within the metric-affine framework \cite{Hehl_McCrea_Mielke_Neeman1995}, the geometrical quantities $g_{ab},\bh^a,\bomega^{a}_{~b}$ are initially independent with 10, 16, and 64 independent elements, respectively. We will assume that our geometries are invariant under the group of  linear transformations of the frame, $GL(4,\mathbb{R})$ which will cause $g_{ab}$ to be dependent on $\bh^a$.  We will also assume that the connection is compatible with the metric.  These two assumptions impose constraints on the geometrical framework. 
\subsection{Gauge choices}

We will refer to the choice of basis for the tangent space as a choice of gauge. This can be accomplished using the $GL(4, \mathbb{R})$ gauge and we can use this freedom in the tangent space to diagonalize the symmetric metric, $g_{ab}= \eta_{ab} = \mbox{Diag}[-1,1,1,1]$, with respect to the coframe, $\{ \bh^a \} = \{ {\bf u}, {\bf x}, {\bf y}, {\bf z}\}$; such a gauge is called the {\bf Orthonormal gauge}.  There is another choice of gauge that is beneficial when working with null directions; in this gauge the coframe is now a null coframe $ \{ \bk^a \} = \{ \bn, \bell, {\bf y}, {\bf z} \}$ and this can be built from the orthonormal frame $\{ {\bf u}, {\bf x}, {\bf y}, {\bf z} \}$ through the transformation
\beq \begin{aligned} & \bn = \frac{1}{\sqrt{2}} ( {\bf u} + {\bf x}),~~ \bell = \frac{1}{\sqrt{2}} ( {\bf u} - {\bf x}).
\end{aligned} \label{orthoframe} \eeq
\noindent The tangent space metric becomes, $$g_{ab} = \eta_{ab} =  \left[ \begin{array}{cccc} 0 & -1 & 0 & 0 \\ -1 & 0 & 0 & 0 \\ 0 & 0 & 1 & 0 \\ 0 & 0 & 0 & 1 \end{array}\right]. $$

In any gauge, but in particular the orthonormal gauge and the null gauge, there exists a subgroup of $GL(4,\mathbb{R})$ of residual gauge transformations that leave the form of the metric unchanged which is known as the orthogonal group, $O(1,3)$, which we restrict further to the special orthogonal subgroup $SO(1,3)$. In this null gauge, the tangent space metric has been completely fixed. The independent dynamical variables arise as the components of the frame and the connection giving 16+64=80 independent elements, albeit with a remaining 6 dimensional $SO(1,3)$ gauge freedom. We will now impose the assumption that the connection be metric compatible, i.e., $Q_{abc}\equiv  - \nabla_{c} g_{ab} =0$, so that the connection becomes anti-symmetric, $\bomega_{(ab)}=0$.

The fundamental variables remaining are the 16 elements of the coframe $\bh^a$ and the 24 elements of the anti-symmetric connection $\bomega^a_{~b}$ while still maintaining the 6 dimensional $SO(1,3)$ gauge freedom.  The torsion and the curvature associated with the coframe and connection are
\beq \begin{aligned} T^a_{~bc} &= \omega^a_{~cb} - \omega^a_{~bc} - c^a_{~bc}, \\
R^a_{~bcd} & = \bh_c [\omega^a_{~bd}] - \bh_d [ \omega^a_{~bc}]+ \omega^e_{~bd} \omega^a_{~ec} - \omega^e_{~bc}\omega^a_{~ed}- c^e_{~cd} \omega^a_{~be}, \end{aligned} \eeq

\noindent where $[\bh_c, \bh_d] = c^e_{~cd} \bh_e$ and $c^e_{~cd}$ are the coefficients of anholonomy for the frame $\bh_a$. 

\subsection{Decomposition of the Curvature tensor}

The curvature tensor can be decomposed into six irreducible parts \cite{obukhov2023poincare, Hehl_McCrea_Mielke_Neeman1995}. The Ricci tensor and co-Ricci tensor are defined as
\beq \begin{aligned}
	 R_{ab} &= R^e_{~aeb}, \\
	 \bar{R}_{ab} &= \frac12 R_{eacd} \epsilon^{cde}_{~~~b},
\end{aligned} \label{Ric} \eeq
\noindent  and $\epsilon^{abcd}$ is the Levi-Civita tensor associated with the Hodge dual. The Ricci scalar and the Ricci pseudo-scalar are then the contractions: 
\beq \begin{aligned}
	R = R^a_{~a}, \\
	\bar{R} = \bar{R}^a_{~a}.
\end{aligned} \label{Rscalar}  \eeq 
\noindent Denoting ${_\mathfrak{s}}T_{ab} = T_{(ab)}$ and ${_\mathfrak{a}} T_{ab} = T_{[ab]}$, respectively, as the symmetric and anti-symmetric parts of an arbitrary rank two tensor, ${\bf T}$. We note that ${_a}\bar{R}^{ab} = {_a}R_{cd} e^{cdab}$. Now looking at the symmetric parts of the Ricci tensor and the co-Ricci tensor, we may construct the trace-free Ricci tensor and the trace-free co-Ricci tensor:
\beq \begin{aligned} 
	S_{ab} &= {_\mathfrak{s}} R_{ab}-\frac14 R g_{ab}, \\
	\bar{S}_{ab} &= {_\mathfrak{s}} \bar{R}_{ab}-\frac14 \bar{R} g_{ab}.
\end{aligned} \eeq
Finally, the Weyl tensor may be written as 
\beq C_{ab}^{~~cd} = R_{ab}^{~~cd} - \bar{S}_{e}^{~[c}\epsilon^{d]e}_{~~~ab} + \frac{1}{12} \bar{R} \epsilon_{ab}^{~~cd} + 2 S_{[a}^{~~[c} \delta^{d]}_{~~b]}+2{_\mathfrak{a}} R_{[a}^{~~[c} \delta^{d]}_{~~b]}+ \frac{1}{6} R \delta_{[a}^{~~[c} \delta^{d]}_{~~b]} . \label{Weyl} \eeq

\noindent To determine the Riemann-Cartan geometries which admit a seven-dimensional symmetry group, we will mainly work with the Weyl tensor, the Ricci tensor and the co-Ricci tensor. 

\subsection{Decomposition of the torsion tensor}

The torsion tensor can be decomposed into three irreducible parts \cite{Hehl_McCrea_Mielke_Neeman1995}: 
\beq T_{abc} = \frac23 (t_{abc} - t_{acb}) - \frac13 (g_{ab} V_c - g_{ac} V_b) + \epsilon_{abcd} A^d.\label{TorsionDecomp} \eeq

\noindent Here ${\bf V}$ denotes the trace of the torsion tensor: 
\beq V_a = T^b_{~ba}. \label{Vtor} \eeq
\noindent Using the Hodge dual of $T_{abc}$, we determine another one-form, ${\bf A}$: 
\beq A^a = \frac16 \epsilon^{abcd}T_{bcd}. \label{Ator} \eeq
\noindent Finally, we can construct the purely tensorial part, ${\bf t}$: 
\beq t_{abc} = \frac12 (T_{abc}+ T_{bac}) -\frac16 (g_{ca} V_b + g_{cb} V_a) + \frac13 g_{ab} V_c. \label{Ttor} \eeq
\noindent The tensors, $V_a$, $A^a$ and $t_{abc}$ are called, respectively, the {\it vector part, axial part, and tensor part } of the torsion tensor. The tensor part satisfies the following identities: 
\beq \begin{aligned}
& g^{ab} t_{abc} = 0, ~t_{abc} = t_{bac},~t_{abc} + t_{bca} + t_{cab} = 0. \end{aligned} \eeq

\noindent A simple counting argument shows that the tensor part of the torsion will have 16 algebraically independent components. Including the four components of the vector part and of the axial part, this yields the 24 components of the torsion tensor.

%-------------------------------------------------------------------------
%%----------------------   SECTION       ----------------------------------
%%-------------------------------------------------------------------------

\section{The Cartan-Karlhede algorithm} \label{sec:CKalg}

The set of components of the curvature tensor and its covariant derivatives up to order $q$ will be denoted as $\mathcal{R}^q$. Similarly, the set of components of the torsion tensor and its covariant derivatives up to order $q$ will be written as $\mathcal{T}^q$.  The Cartan-Karlhede algorithm has been discussed in detail in \cite{Coley:2019zld, fonseca1996algebraic} and we will only briefly review the algorithm here.

The zeroth iteration of the CK algorithm begins by computing the curvature tensor and torsion tensor. Using canonical forms for these tensors the frame basis is adapted to realize the canonical form for one or both tensors. The subgroup of the Lorentz frame transformations that do not change the chosen form for the tensors is determined. This is called the zeroth order linear isotropy group, $H_0$. The number of functionally independent components in $\mathcal{R}^0$ and $\mathcal{T}^0$ is also recorded and denoted as $t_0$. 

The algorithm always continues to the first iteration. For the $q^{th}$ iteration, $q > 0$, the $q^{th}$ covariant derivatives of the curvature tensor and torsion tensor are computed. Using the $(q-1)^{th}$ linear isotropy group, determine the canonical form for the $q^{th}$ covariant derivatives of the curvature tensor and torsion tensor, and fix the frame accordingly. Then record the $q^{th}$ linear isotropy group, $H_q$ which leaves the $q^{th}$ covariant derivatives of the curvature tensor and torsion tensor unchanged. Similarly, record the number of functionally independent components,  $t_q$, in $\mathcal{R}^q$ and $\mathcal{T}^q$. 

The algorithm provides two helpful sequences of discrete invariants, $dim~H_{q}$ and $t_q$ for $q=0,\ldots, p+1$ which provides a stopping condition for the algorithm. The stopping condition occurs at some iteration, $p+1$, when the dimension of the linear isotropy group and the number of functionally independent invariants are the same as at iteration $p$. That is, $dim~H_p = dim~H_{p+1}$ and $t_p = t_{p+1}$. The resulting non-zero components of $\mathcal{R}^{p+1}$  and $\mathcal{T}^{p+1}$ constitute the set of {\it Cartan invariants} and we will denote them as $\mathcal{R} \equiv \mathcal{R}^{p+1}$ and $\mathcal{T} \equiv \mathcal{T}^{p+1}$. 

At the conclusion of the algorithm, we have determined a geometrically preferred frame which can be used to characterize a given Riemann-Cartan geometry. If $dim(H_p) \neq 0$, the frame arising from the CK algorithm is known as an {\it invariantly defined frame fixed completely up to linear isotropy}. While if $H_p$ is the trivial group, so that $dim(H_p) =0$, the frame is called an {\it invariant frame}. For sufficiently smooth frames and connections, the 4D spacetime is uniquely locally characterized by the canonical forms of the curvature tensor, torsion tensor and their covariant derivatives, the two discrete sequences arising from the successive linear isotropy groups and the independent function counts, and the values of the (non-zero) Cartan invariants. 

As there are $t_p$ essential coordinates, $x^{\alpha'}$, the remaining $4-t_p$ coordinates $x^{\tilde{\alpha}}$ are ignorable, and so the dimension of the affine frame symmetry isotropy group (hereafter called the isotropy group) of the spacetime will be $s=\dim(H_p)$ and the affine frame symmetry group has dimension: \beq r=s+4-t_p. \label{rnumber} \eeq

\subsubsection{Conditions for maximal symmetry}
 
For any torsionful or non-flat Riemann Cartan geometry to achieve the largest affine frame symmetry group, $t_p$ should be zero, implying that all Cartan invariants are constant, and hence the geometry is locally homogeneous and $s$ should be as large as possible. In the case of Riemann-Cartan geometries which are not equivalent to the Minkowski geometry, the torsion tensor is non-vanishing and this puts a limit on the linear isotropy group, $H_p$. It was shown in \cite{Coley:2019zld} that the largest isotropy group permitted occurs when the purely tensorial part of the torsion tensor vanishes \beq t_{abc} = 0 \label{TTor3Iso} \eeq
\noindent  and the vector part, $V^a$, and axial part, $A^a$ are proportional to each other:
 \beq V^a \propto A^a. \label{AV3Iso} \eeq 
\noindent In each case, we will adapt the frame so that $V^a$ and $A^a$ are a constant multiple of a frame basis element. 

This gives three possibilities depending on whether $V^a$ is timelike, spacelike or null. Denoting the norm of $V^a$ as $|V|^2 = g_{ab} V^a V^b$, the vector-field is timelike if $|V|^2 <0$, spacelike if $|V|^2>0$ and null if $|V|^2=0$. Furthermore, we will fix the frame so that ${\bf V} = U_0 \bh^1$ when $|V|^2<0$, ${\bf V} = U_0 \bh^2$ when $|V|^2>0$ and ${\bf V} = U_0 \bk^1$ when $|V|^2=0$, where $U_0$ is a constant. The corresponding linear isotropy groups are the stabilizer of the invariant direction. These are $SO(3)$ for timelike vector-fields, $SO(1,2)$ for spacelike vector-fields and $E(2)$ for null vector fields. 

We have only considered the torsion tensor so far. Requiring that the Riemann tensor shares the same linear isotropy group, the Ricci tensor and the co-Ricci tensor must satisfy the following form for the timelike and spacelike cases 
\beq 
\begin{aligned}
R_{ab} &= R_1 (g_{ab} - |V|^{-2} V_a V_b ) + R_2 |V|^{-2} V_a V_b, \\
\bar{R}_{ab} &= R_3 (g_{ab} - |V|^{-2} V_a V_b ) + R_4 |V|^{-2}  V_a V_b  ,
\end{aligned}  \label{Ric3Iso} 
\eeq
\noindent or in the null case
\beq
\begin{aligned} R_{ab} &= R_2 V_a V_b, \\
\bar{R}_{ab} &= R_4 V_a V_b.
\end{aligned} \label{Ric3IsoNull} 
\eeq
\noindent In addition, the Weyl tensor must vanish \cite{kramer}, 
 \beq C_{abcd} = 0. \label{Weyl3Iso} \eeq 
\noindent Covariant derivatives of the curvature tensor and torsion tensor must share the same linear isotropy and this will lead to further conditions on the vierbein and connection components. In particular, the covariant derivative of the vector part of the torsion must take the form 
\beq V_{a;b} = U_1 (g_{ab} - |V|^{-2} V_a V_b ) + U_2 |V|^{-2}  V_a V_b \label{DV3IsoST} \eeq
\noindent if $|V|^2 \neq 0$ and if $|V|^2 =0$ then 
\beq V_{a;b} =  U_2 V_a V_b. \label{DV3IsoNull} \eeq

\noindent This ensures that all higher order covariant derivatives admit the same maximal linear isotropy group. We note that $U_0, R_1, R_2, R_3, R_4, U_1$ and $U_2$ must be constants to ensure the largest symmetry group dimension given by equation \eqref{rnumber}.
\\

\subsubsection{Lorentz frame transformations}
The Lorentz frame transformations are most easily summarized in a null coframe $\{ \bell, \bn, \bm, \bar{\bm}\}$ where \beq \bm = \frac{{\bf y}+i {\bf z}}{\sqrt{2}},~~\bar{\bm} = \frac{{\bf y}-i {\bf z}}{\sqrt{2}}. \label{yz_to_m} \eeq

\noindent The frame transformations are then: \\

\begin{itemize}

\item Boosts and spins with real-valued parameters $B$ and $\Theta$, respectively: 

\beq \bell' = B \bell,~~\bn' = \frac{\bn}{B},~ \bm' = e^{i \Theta} \bm,~ \bar{\bm}' = e^{-i\Theta} \bar{\bm}. \label{null_boost_spins} \eeq

\item Null rotations about $\bell$ with complex-valued parameter $L$:

\beq \bell' = \bell,~ \bn'  = \bn + L \bm + \bar{L} \bar{\bm} + |L|^2 \bell, ~ \bm' = \bm + \bar{L} \bell \label{null_l_rotate} \eeq

\item Null rotations about $\bn$ with complex-valued parameter $N$:

\beq \bell' = \bell + N \bm + \bar{N} \bar{\bm} + |N|^2 \bn,~ \bn'  = \bn',~ \bm' = \bm + \bar{N} \bn. \label{null_n_rotate} \eeq

\end{itemize}

These transformations can be constructed in an orthonormal frame by combining boosts in planes spanned by the timelike direction and a spacelike direction, with spatial rotations in planes spanned by two spacelike directions. 

The boost and rotational frame transformations will be discussed using two examples which are generalized in a straightforward manner to any pair of basis elements. Given an orthonormal frame, if we consider a boost in the ${\bf u} - {\bf x}$ plane, with real-valued parameter $B$, the resulting transformation is then
\beq \begin{aligned} & {\bf u}' = \frac{B^2+1}{2B} {\bf u} + \frac{B^2-1}{2B} {\bf x},\\
& {\bf x}' = \frac{B^2-1}{2B} {\bf u} + \frac{B^2+1}{2B}  {\bf x}, \\
& {\bf y} ' = {\bf y},~ {\bf z}' = {\bf z}.
\end{aligned} \label{Ortho_Boost}  \eeq

\noindent A rotation in the ${\bf y}-{\bf z}$ plane, with parameter $\Theta$ is then

\beq \begin{aligned} & {\bf u'} = {\bf u},~ {\bf x}' = {\bf x'} \\
& {\bf y}' = \cos(\Theta) {\bf y} + \sin(\Theta) {\bf z}, \\
& {\bf z}' = -\sin(\Theta) {\bf y} + \cos(\Theta) {\bf z}. \label{Ortho_Rotate} \end{aligned} \eeq

%-------------------------------------------------------------------------
%%----------------------   SECTION       ----------------------------------
%%-------------------------------------------------------------------------

\section{Affine frame symmetries with isotropy} \label{sec:FSs}

In the previous section, we saw that the CK algorithm is able to determine the number of affine frame symmetries for a given Riemann-Cartan geometry. Using the resulting invariantly defined frame basis, $\bh^a$, determined by the CK algorithm, we can also determine conditions for an affine frame symmetry. If at the conclusion of the algorithm, the linear isotropy group is non-trivial, $dim~H_p \neq 0$, then the infinitesimal action of an affine frame symmetry $\phi_{\tau}$ with parameter $\tau$ is then \cite{olver1995equivalence}:
\beq \mathcal{L}_{{\bf X}} \bh^a = \lambda^a_{~b} \bh^b, \label{Liederivative:frame} \eeq

\noindent where ${\bf X}$ is the generator of the affine frame symmetry, and  $\lambda^a_{~b}$ is the Lie algebra generator for $\Lambda^a_{(\tau)~b} \in H_p$ associated with $\phi_{\tau}$. 

This condition is not sufficient to define ${\bf X}$ as an affine frame symmetry generator. Strictly speaking, an affine frame symmetry must leave the curvature tensor, the torsion tensor, and their covariant derivatives unchanged. To satisfy this condition, we impose that ${\bf X}$ is an affine collineation \cite{aaman1998riemann}: 
\beq (\mathcal{L}_{{\bf X}} \bomega)^a_{~bc} =0. \label{Liederivative:Con} \eeq
\noindent Using the coordinate basis expression for the Lie derivative of the connection in  \cite{yano2020theory,lichnerowicz1977geometry, Trautman:2006fp}, the corresponding frame basis expression is:

\beq (\mathcal{L}_{{\bf X}}\bomega)^a_{~bc} \bh_a = {\bf R}( {\bf X},\bh_c) \bh_b - {\bf T}({\bf X}, \nabla_c \bh_b) + \nabla_c {\bf T}({\bf X}, \bh_b) + \nabla_c \nabla_b {\bf X} - \nabla_{\nabla_c \bh_b} {\bf X}, \label{eq:RC:AC} \eeq

\noindent where ${\bf R}(\bh_d, \bh_c)\bh_b = R^a_{~bcd} \bh_a$ and ${\bf T} ( \bh_c, \bh_b) = T^a_{~bc} \bh_a$, respectively, denote the curvature tensor and torsion tensor of the connection $\omega^a_{~bc}$.

It follows from the definition of the curvature tensor and the torsion tensor that requiring equations  \eqref{Liederivative:frame} and \eqref{Liederivative:Con} to hold is sufficient for the Lie derivative with respect to ${\bf X}$ of the curvature tensor and the torsion tensor to be zero and  it follows that the Lie derivative with respect to ${\bf X}$ of their covariant derivatives are also zero. 

The above characterization of affine frame symmetries is not helpful for generating Riemann-Cartan geometries that permit a given group of symmetries. We must know the form of the curvature tensor, the torsion tensor and their covariant derivatives in order to determine an invariantly defined frame, which presupposes knowledge of the Riemann-Cartan geometry. To avoid this practical issue, a more general class of frames need to be introduced. They are a generalization of the invariantly defined frames defined by the CK algorithm in the sense that under Lie differentiation with an affine frame symmetry generator, the new frames are influened by the Lie algebra generators of the isotropy group instead of the linear isotropy group. 
 
\begin{defn} \label{defn:SymFrame}
For a Riemann-Cartan geometry, $({\bh}^a, \bomega^a_{~b})$, which admits an affine frame symmetry, ${\bf X}$, we define the class of {\bf symmetry frames}, ${\bh}^a$, as those frames which satisfy the condition 
\beq && \mathcal{L}_{{\bf X}} \bh^a = f_{X}^{~\hat{i}} \lambda^a_{\hat{i}~b} \bh^b. \label{TP:frm:sym} \eeq

\noindent Here, the functions $f_X^{~\hat{i}}(x^{\alpha'})$ are arbitrary and $\lambda^a_{\hat{i}~b}$ are basis elements of the Lie algebra of the isotropy group. The index $\hat{i}$ goes from 1 to the dimension of the isotropy group. 

\end{defn}

Using this definition, one can show that an affine frame symmetry generator, ${\bf X}$, is necessarily a Killing vector field since equation \eqref{TP:frm:sym} and the definition of a frame basis implies that the Lie derivative of the metric must vanish. We note that the invariantly defined frame arising from the CK algorithm is a symmetry frame but that not all symmetry frames are invariantly defined. 

A new subgroup of the Lorentz frame transformations, $\overline{Iso}$, is defined to be the group of Lorentz frame transformations that leaves the representation of the isotropy group unchanged \cite{McNutt:2023nxmnxm}. Using $\overline{Iso}$, we can transform any symmetry frame to a new symmetry frame. Correspondingly, the functions $f_X^{~\hat{i}}$ transform under a frame transformation, $\tilde{\bh}^a = \tilde{\Lambda}^a_{~b} \bh^b$ where $\tilde{\Lambda}^a_{~b} \in \overline{Iso}$ as: 
\beq {\bf X}_I ( \tilde{\Lambda}^a_{~b}) [\tilde{\Lambda}^{-1}]^b_{~c} + \tilde{\Lambda}^a_{~b} f_I^{~\hat{i}} \lambda^b_{\hat{i}~d} [\tilde{\Lambda}^{-1}]^d_{~c} = \tilde{f}_I^{~\hat{i}} \lambda^a_{\hat{i}~ c}. \label{frot} \eeq
 
Despite this transformation rule, the functions $f_X^{~\hat{i}}$ are tensor quantities since they appear in equation \eqref{TP:frm:sym}. We may use the parameters of the Lorentz frame transformations in $\overline{Iso}$ to fix the form of $f_X^{~\hat{i}}$ in an invariant manner, such as by setting some of $f_X^{~\hat{i}}$ to be constant or zero. By doing this, we restrict the frame freedom to a smaller subgroup, $\overline{H}_p \subset	\overline{Iso}$, of Lorentz frame transformations that leaves the chosen form of $f^{~\hat{i}}_{X}$ unchanged. If  after the functions $f_{ X}^{~\hat{i}}$ are fixed as much as possible with a non-trivial isotropy group $\overline{H}_p$, then the symmetry frame is an {\it invariantly defined symmetry frame up to the isotropy group $\overline{H}_p$}. However, if $\overline{H}_p$ consists of only the identity Lorentz frame transformation, then the symmetry frame is now an {\it invariant symmetry frame}. 

The initial definition of an affine frame symmetry can be restated in terms of a symmetry frame. 
\begin{defn} \label{AFS2}
An {\it affine frame symmetry} generator is a vector field satisfying
\beq &&\mathcal{L}_{{\bf X}_I} \bh^a = f_I^{~\hat{i}} \lambda^a_{~\hat{i}~b} \bh^b \label{TP:sym1} \\ 
&& (\mathcal{L}_{{\bf X}_I} \bomega)^a_{~bc} = 0 , \label{TP:sym2} \eeq

where $\hat{i}, \hat{j}, \hat{k} \in \{1, \ldots, n\}$ and the functions, $f_I^{~\hat{i}}$, depend on the coordinates, $x^{\alpha'}$. 
\end{defn}

This definition has the advantage that we do not require prior knowledge of the curvature tensor, the torsion tensor, or their covariant derivatives to specify an affine frame symmetry. However, this definition is only sufficient to determine all possible Riemann-Cartan geometries that admit a single affine frame symmetry. If we are interested in Riemann-Cartan geometries with a larger group of affine frame symmetries, there are further conditions that must be specified. 

Supposing that our Riemann-Cartan geometry admits a Lie group of affine frame symmetries of dimension $N$, with a non-trivial isotropy group with dimension $n$ $(n < N)$, we may choose coordinates so that the affine frame symmetry group is represented as a set of vector fields, ${\bf X}_I$, where $I, J, K \in \{1, \ldots, N\}$, with the Lie algebra commutator relations: 
\beq [{\bf X}_I, {\bf X}_J] = C^K_{~IJ} {\bf X}_K, \label{CijkSym} \eeq

\noindent where the coefficients, $C^K_{~IJ}$, are structure constants of the Lie algebra. To determine all possible Riemann-Cartan geometries which admit a given Lie group as as a group of affine frame symmetries, we must find an orthonormal symmetry coframe $\bh^a$, with a metric-compatible connection, $\omega^a_{~bc}$ satisfying the conditions in definition \eqref{AFS2}. We note that a null symmetry coframe, $\bk^a$, is also applicable when the affine symmetry group warrants this choice. 

In equation \eqref{TP:sym1} we must consider an array of functions, $f_I^{~\hat{i}}$, where the index $I$ can be $1$ to $N$, and these can be invariantly specified as before.  This process yields a class of invariantly defined symmetry frames up to the linear isotropy in $\overline{H}_q$. For a general symmetry frame, the properties of the Lie derivative and equation \eqref{CijkSym} gives a new condition on the functions $f_I^{~\hat{i}}$:
\beq [\mathcal{L}_{{\bf X}_I}, \mathcal{L}_{{\bf X}_J}] \bh^a = \mathcal{L}_{[{\bf X}_I, {\bf X}_J] } \bh^a = C^K_{~IJ} \mathcal{L}_{{\bf X}_K} \bh^a. \label{TP:sym3} \eeq
\noindent Then writing the associated Lie algebra of the isotropy group as: \beq [\lambda_{\hat{i}}, \lambda_{\hat{j}}]^a_{~b} = \lambda^a_{\hat{i}~b} \lambda^b_{\hat{j}~c} - \lambda^a_{\hat{j}~b} \lambda^b_{\hat{i}~c} =  C^{\hat{k}}_{~\hat{i} \hat{j}} \lambda^a_{\hat{k}~c}. \nonumber \eeq

\noindent equations \eqref{TP:sym1}, \eqref{TP:sym3} and \eqref{eq:RC:AC} can be restated in the following proposition.

\begin{thm} \label{Sym:RC:Prop}
The most general Riemann-Cartan geometry admitting a given group of affine frame symmetries having vector generators, ${\bf X}_I,~ I,J,K \in \{1, \ldots, N\}$ with an isotropy group of dimension $n$ can be determined by solving for the unknowns $h^a_{~\mu}$, $f_I^{~\hat{i}}$ (with $\hat{i}, \hat{j}, \hat{k} \in \{1,\ldots, n\}$) and $\omega^a_{~bc}$ from the following equations:
\beq 
& { X}_I^{~\nu} \partial_{\nu} h^a_{~\mu} + \partial_{\mu} { X}_I^{~\nu} h^a_{~\nu} = f_I^{~\hat{i}} \lambda^a_{\hat{i}~b} h^b_{~\mu}, & \label{TP:sym1big} \\
& 2{\bf X}_{[I} ( f_{J]}^{~\hat{k}}) - f_I^{~\hat{i}} f_J^{~\hat{j}} C^{\hat{k}}_{~\hat{i} \hat{j}} = C^K_{~IJ} f_K^{~\hat{k}}, & \label{TP:sym3big}\\
& { X}_I^{~d} \bh_d( \omega^a_{~bc}) + \omega^d_{~bc} f_I^{~\hat{i}} \lambda^a_{\hat{i}~d} - \omega^a_{~dc} f_I^{~\hat{i}} \lambda^d_{\hat{i}~b} - \omega^a_{~bd} f_I^{~\hat{i}} \lambda^d_{\hat{i}~c} - \bh_c( f_I^{~\hat{i}}) \lambda^a_{\hat{i}~b} = 0, \label{SFcon} & \eeq

\noindent where $\{ \lambda^a_{\hat{i}~b}\}_{\hat{i}=1}^n$ is a chosen basis for the Lie algebra of the  isotropy group, $[\lambda_{\hat{i}}, \lambda_{\hat{j}}] = C^{\hat{k}}_{~\hat{i}\hat{j}} \lambda_{\hat{k}}$, $[{\bf X}_I, {\bf X}_J] = C^K_{~IJ} {\bf X}_K$. 

\end{thm}

This proposition applies for any symmetry frame but the solution to these equations is most readily found using an invariantly defined symmetry frame up to $\overline{H}_q$, where the number of free functions, $f_I^{~\hat{i}}$ has been reduced as much as possible.

%-------------------------------------------------------------------------
%%----------------------   SECTION       ----------------------------------
%%-------------------------------------------------------------------------

\section{Riemann-Cartan geometries with 3-isotropy} \label{sec:3isospaces}

The geometries that can potentially admit maximal linear isotropy can be identified by considering the subgroups of the Lorentz group that leave a direction invariant. There are three cases, timelike, spacelike and null. The corresponding isotropy groups are isomorphic to $SO(3), SO(1,2)$ or $E(2)$. Choosing a coordinate system for each of these groups we have

\begin{itemize}
\item $SO(3)$: $\{ t, r, \theta, \phi\}$
\beq {\bf X}_1 = \sin(\phi) \partial_\theta + \frac{\cos(\phi)}{\tan(\theta)} \partial_\phi,~{\bf X}_2 = - \cos(\phi) \partial_\theta + \frac{\sin(\phi)}{\tan(\theta)} \partial_\phi,~{\bf X}_3 = \partial_\phi. \eeq

\noindent With the commutator relations:
\beq [{\bf X}_1, {\bf X}_2] = -{\bf X}_3,~[{\bf X}_1, {\bf X}_3] = {\bf X}_2,~[{\bf X}_2, {\bf X}_3] = -{\bf X}_1. \eeq 

\item $SO(1,2)$: $ \{t,\theta, \phi, z\}$
\beq {\bf X}_1 = - \cos(\phi) \partial_\theta + \frac{\sin(\phi)}{\tanh(\theta)} \partial_\phi,~{\bf X}_2 = \sin(\phi) \partial_\theta + \frac{\cos(\phi)}{\tanh(\theta)} \partial_\phi,~{\bf X}_3 = \partial_\phi.  \eeq 

\noindent With the commutator relations:
\beq [{\bf X}_1, {\bf X}_2] = -{\bf X}_3,~[{\bf X}_1, {\bf X}_3] = -{\bf X}_2,~[{\bf X}_2, {\bf X}_3] = {\bf X}_1. \eeq

\item $E(2)$: $\{u,v, r, \theta\}$
\beq {\bf X}_1 = - \cos(\theta) \partial_r + \frac{\sin(\theta)}{r} \partial_\theta,~{\bf X}_2 = \sin(\theta) \partial_r + \frac{\cos(\theta)}{r} \partial_\theta,~ {\bf X}_3 = \partial_\theta. \eeq

\noindent With the commutator relations:
\beq [{\bf X}_1, {\bf X}_2] = 0,~[{\bf X}_1, {\bf X}_3] = -{\bf X}_2,~[{\bf X}_2, {\bf X}_3] = {\bf X}_1. \eeq

\end{itemize}

\noindent We note that in the last case, we have chosen a coordinate system where a singularity occurs at $r=0$. This can always be removed by choosing new coordinates, $\tilde{x} = r \cos(\theta)$ and $\tilde{y} = r \sin(\theta)$. 

Finally to employ equation \eqref{TP:sym1big}, we must specify a representation of our isotropy subalgebra. The representation for $\lambda^a_{\hat{i}~b}$ are then 
\begin{itemize}
\item $SO(3)$:
\beq \lambda_{\hat{1}} = \left[ \begin{array}{cccc} 0 & 0 & 0 & 0\\ 0 & 0 & 0 & 0 \\ 0 & 0 & 0 & 1 \\ 0 & 0 & -1 & 0 \end{array}\right],~ \lambda_{\hat{2}} = - \left[ \begin{array}{cccc} 0 & 0 & 0 & 0\\ 0 & 0 & 1 & 0 \\ 0 & -1 & 0 & 0 \\ 0 & 0 & 0 & 0 \end{array}\right],~ \lambda_{\hat{3}} = -  \left[ \begin{array}{cccc} 0 & 0 & 0 & 0\\ 0 & 0 & 0 & 1 \\ 0 & 0 & 0 & 0 \\ 0 & -1 & 0 & 0 \end{array}\right]. \label{so3:iso} \eeq
\item $SO(1,2)$:
\beq \lambda_{\hat{1}} = \left[ \begin{array}{cccc} 0 & 0 & 1 & 0\\ 0 & 0 & 0 & 0 \\ 1 & 0 & 0 & 0 \\ 0 & 0 & 0 & 0 \end{array}\right],~ \lambda_{\hat{2}} = \left[ \begin{array}{cccc} 0 & 1 & 0 & 0\\ 1 & 0 & 0 & 0 \\ 0 & 0 & 0 & 0 \\ 0 & 0 & 0 & 0 \end{array}\right],~ \lambda_{\hat{3}} = \left[ \begin{array}{cccc} 0 & 0 & 0 & 0\\ 0 & 0 & 1 & 0 \\ 0 & -1 & 0 & 0 \\ 0 & 0 & 0 & 0 \end{array}\right]. \label{so12:iso} \eeq
\item $E(2)$:
\beq \lambda_{\hat{1}} = -\left[ \begin{array}{cccc} 0 & 0 & 0 & 0\\ 0 & 0 & 0 & 1 \\ 0 & 0 & 0 & 0 \\ 1 & 0 & 0 & 0 \end{array}\right],~ \lambda_{\hat{2}} = -\left[ \begin{array}{cccc} 0 & 0 & 0 & 0\\ 0 & 0 & 1 & 0 \\ 1 & 0 & 0 & 0 \\ 0 & 0 & 0 & 0 \end{array}\right],~ \lambda_{\hat{3}} = \left[ \begin{array}{cccc} 0 & 0 & 0 & 0\\ 0 & 0 & 0 & 0 \\ 0 & 0 & 0 & 1 \\ 0 & 0 & -1 & 0 \end{array}\right]. \label{e2:iso} \eeq

\end{itemize}

\noindent In the last case, this representation necessitates the use of a null coframe $\{ \bk^a \}$. 

With these symmetry groups and their associated representations defined, we will determine a symmetry frame by specifying the form of $f_I^{~\hat{i}}$ in equation \eqref{TP:sym1big} using equation \eqref{TP:sym3big}. The approach is similar to the spherically symmetric case in \cite{McNutt:2023nxmnxm} and yields the following non-zero components

\begin{itemize}
\item $SO(3)$
\beq f_I^{~\hat{i}} =  \left[ \begin{array}{ccc} \frac{\cos(\phi)}{\sin(\theta)} & 0 & 0 \\
\frac{\sin(\phi)}{\sin(\theta)} & 0 & 0 \\
0 & 0 & 0 \end{array} \right] \eeq

\item $SO(1,2)$
\beq f_I^{~\hat{i}} =  \left[ \begin{array}{ccc} 0 & 0 & \frac{\sin(\phi)}{\sinh(\theta)}  \\
0 & 0 & \frac{\cos(\phi)}{\sinh(\theta)} \\
0 & 0 & 0 \end{array} \right] \eeq

\item $E(2)$
\beq f_I^{~\hat{i}} =  \left[ \begin{array}{ccc} 0 & 0 & \frac{\sin(\theta)}{r}  \\
0 & 0 & \frac{\cos(\theta)}{r} \\
0 & 0 & 0 \end{array} \right] \eeq

\end{itemize}
 
Employing the above theorems we find the following three invariantly defined symmetry coframes up to their respective linear isotropy groups. We will omit the expressions for the connections, these are included in the appendix. Instead, we will only write down the associated connections when we consider the $G_4$ analogues of these spaces

\begin{itemize}

\item $SO(3)$: 
\beq \bh^1 = \alpha(t,r) dt,~~\bh^2 = \beta(t,r) dr,~~\bh^3 = \gamma(t,r) d\theta,~~ \bh^4 = \gamma(t,r) \sin(\theta) d\phi.  \label{eq:so3frame} \eeq
 \item $SO(1,2)$: 
 \beq \bh^1 = \alpha(t,z) dt,~~ \bh^2 = \beta(t,z) d\theta,~~ \bh^3 = \beta(t,z) \sinh(\theta) d\phi,~~ \bh^4 = \gamma(t,z) dz. \label{eq:so12frame} \eeq
 \item $E(2)$: 
 \beq \bk^1 = du,~~\bk^2 = \alpha(u,v)du+\beta(u,v)dv,~~ \bk^3 = \gamma(u,v) dr, ~~\bk^4 = \gamma(u,v) d \theta. \label{eq:e2frame} \eeq
\end{itemize}

We note that in the case of $SO(3)$ and $SO(1,2)$ geometries, it is possible to choose coordinates to simplify the vierbein so that either $\gamma(t,r) = r$, if $\gamma$ is not constant, in the case of $SO(3)$ or $\beta(t,z) = z$, if $\beta$ is not constant, in the case of $SO(1,2)$. However, this will only be helpful when considering particular subclasses of the $G_4$ spacetimes. For the $E(2)$ case, the form of the affine frame symmetry restricts the freedom to specify the arbitrary functions in the general case.

\section{Extension to $G_4$ Riemann-Cartan geometries}  \label{sec:G4spaces}

For each of the 3-isotropic spaces, there is a direction that is left invariant by the symmetry generators. By requiring that there exists a fourth symmetry generator, ${\bf Y}$, in this direction, we can further specialize the geometries, to better determine the $G_4$ geometries. 

Following Schmidt's method as outlined in \cite{kramer}, for each of the Lie algebras $\{ {\bf X}_I \}_{I=1}^3$ with structure constants $C^I_{~JK}$ we suppose there is an additional symmetry vector generator ${\bf Y}$ such that 
\beq [{\bf X}_I, {\bf Y}] = D {\bf Y} + F_I^{~J} {\bf X}_J. \eeq

\noindent If ${\bf Y}$ is proportional to the invariant direction, then $D = 0$ and it is straightforward to show that the Jacobi identities are satisfied only if $[{\bf X}_I, {\bf Y}] = 0$ for $I=1,2,3$. 

Introducing a fourth affine frame symmetry vector field, ${\bf Y} = Y^a \bh_a$ we can employ the coordinate freedom in each case to specify the form of ${\bf Y}$ and conditions on an arbitrary function to determine distinct sub-cases:

\begin{itemize}

\item $SO(3)$:  $|{\bf Y}|^2 <0$ 
	\begin{enumerate}
	
	\item ${\bf Y} = -C_0 \partial_t + r \partial_r$, with $C_0 \neq 0$ \\ $\a(t,r) = \a(re^{\frac{t}{C_0}}), \b(t,r) = \b(re^{\frac{t}{C_0}})e^{\frac{t}{C_0}}$ and $\e(t,r) = \e(re^{\frac{t}{C_0}})$. 
	\item ${\bf Y} = \partial_t$,\\
		$\a(t,r) = \a(r), \b(t,r)=\b(r)$ and $\e(t,r)= r$,\\
		or \\
		$\a(t,r) = \a(r), \b(t,r)=\b(r)$ and $\e(t,r)= \e_0$, with $\e_0$ a constant.	
	\end{enumerate}
	
\item $SO(1,2)$:  $|{\bf Y}|^2 >0$
	\begin{enumerate}
	\item ${\bf Y} = t \partial_t + C_0 \partial_z$ with $C_0 \neq 0$ \\ $\a(t,z) = \a(te^{-\frac{z}{C_0}})e^{-\frac{z}{C_0}}, \b(t,z) = \b(te^{-\frac{z}{C_0}})$ and $\e(t,z) = \e(te^{-\frac{z}{C_0}})$.
	\item ${\bf Y} = \partial_z$, \\
	$\a(t,z) = \a(t), \b(t,z)=t$ and $\e(t,r)=\e(t)$, \\ 
	or \\
	$\a(t,z) = \a(t), \b(t,z)=\b_0$ and $\e(t,r)=\e(t)$, with $\b_0$ a constant.	
	\end{enumerate}

\item $E(2)$: $|{\bf Y}|^2 =0$
	\begin{enumerate}
	\item ${\bf Y} = \partial_v$, 
	\\
	 $\a(u,v) = \a(u ), \b(u,v)= \b(u)$, and $\e(u,v)= \e(u)$
	
	\end{enumerate}

\end{itemize}

\begin{rem}
In the null case, there are two options for the null direction $\bk_1 = \partial_v$ or $ \bk_2 = \partial_u - \frac{\a}{\b} \partial_v$. However, the resulting solutions can be shown to be equivalent, up to exchanging $u$ and $v$ by using coordinates to rectify $\bk_2$.  Choosing ${\bf Y} \propto \partial_v$, it is always possible to choose coordinates so that ${\bf Y} = \partial_v$. A final change of the $u$ coordinate permits setting $\a = 1$ as all of the functions must now be independent of the $v$ coordinate.
\end{rem}

To continue the analysis we will rectify the vector field in each case to reduce the problem to solving a system of ordinary differential equations. This permits a compact expression for the connection coefficients regardless of what form the fourth affine frame symmetry generator takes. That is, we will use the symmetry adapted coordinates $r'$ or $t'$ for $SO(3)$ and $SO(1,2)$, respectively, in the remainder of the paper. However, after this section we will omit the primes for the sake of presentation. This choice of coordinates will yield connection one-forms that are applicable regardless of the form of ${\bf Y}$. 

\begin{itemize}
\item $SO(3)$: In new coordinates, $\{ t', r', \theta, \phi\}$ with ${\bf Y} = \partial_{t'}$: 
\beq \begin{aligned}  \bomega_{12} & = W_5(r') \bh^1 + W_6(r') \bh^2, \\
\bomega_{13} & = W_7(r') \bh^3 + W_8(r') \bh^4, \\
\bomega_{14} & = - W_8(r') \bh^3 + W_7(r') \bh^3, \\
\bomega_{23} & = W_3(r') \bh^3 + W_4(r') \bh^4, \\
\bomega_{24} & = - W_4(r') \bh^3 + W_3(r') \bh^4, \\
\bomega_{34} & = W_1(r') \bh^1 + W_2(r') \bh^2 - \frac{\cot(\theta)}{\e(r')} \bh^4. \label{SO3connection} \end{aligned} \eeq

\item $SO(1,2)$: In new coordinates $\{ t', \theta, \phi, z'\}$ with ${\bf Y} = \partial_{z'}$ 
\beq \begin{aligned} \bomega_{12} &= W_7(t') \bh^2 +W_8(t') \bh^3, \\
\bomega_{13} & = - W_8(t') \bh^2 + W_7(t') \bh^3, \\
\bomega_{14} & = W_5(t') \bh^1 + W_6(t') \bh^4, \\
\bomega_{23} & = W_1(t') \bh^1 - \frac{\coth(\theta)}{\b(t')} \bh^3 + W_2(t') \bh^4, \\
\bomega_{24} & = W_3(t') \bh^2 + W_4(t') \bh^3, \\
\bomega_{34} & = -W_4(t') \bh^2 + W_3(t') \bh^3. \label{SO12connection} \end{aligned} \eeq

\item $E(2)$: In new coordinates $\{ u', v', r,\theta\}$ with ${\bf Y} = \partial_{v'}$:
\beq \begin{aligned} \bomega_{12} & = W_3(u') \bk^1 + W_8(u') \bk^2, \\
\bomega_{13} & = W_4(u') \bk^3 + W_7(u') \bk^4, \\
\bomega_{14} & = -W_7(u') \bk^3 + W_4(u') \bk^4, \\
\bomega_{23} & = W_6 (u') \bk^3 + W_5(u') \bk^4, \\
\bomega_{24} & = -W_5(u') \bk^3 +W_6(u') \bk^4, \\
\bomega_{34} & = W_1(u') \bk^1 + W_2(u') \bk^2 - \frac{1}{\e(u') r} \bk^4. \label{E2connection} \end{aligned} \eeq

\end{itemize}

\section{Locally homogeneous maximally isotropic Riemann-Cartan Geometries} \label{sec:G7analysis}

For the sake of completeness, we will briefly review the spherically symmetric case. This was fully analysed in \cite{McNutt2024} where static and stationary spherically symmetric geometries were investigated for potential $G_7$ solutions. We will then present the analysis for the cases where the linear isotropy group is either $SO(1,2)$ or $E(2)$ in greater detail. 

\subsection{$SO(3)$ isotropy} \label{sec:SO3}

We will consider the two distinct forms of the fourth affine frame symmetry as separate cases. 

\subsubsection{Case 1} \label{subsec:SO3 Case 1}
In this case, we may always rescale $t$ to set $C_0 = 1$, then the vierbein relative to the symmetry adapted coordinates $\{t, r, \theta, \phi \}$ is:
\beq h^a_{~\mu} = \left[ \begin{array}{cccc} \a(r) & 0 & 0 & 0 \\ 
-r \b(r)  & \b(r) & 0 & 0 \\ 0 & 0 & \e(r) & 0 \\ 0 & 0 & 0 & \e(r) \sin \theta \end{array} \right]. \label{SO3_VB_c1}  \eeq

\noindent Using this frame and the connection in equation \eqref{SO3connection}, we may compute the curvature tensor and torsion tensor. Following from the discussion in section \ref{sec:CKalg}, we require that the tensor-part of the torsion vanishes, which gives the following conditions:
\beq W_4 = W_2, W_8=W_1,  W_5 = W_3 + \frac{\e_{,r}}{\b \e}- \frac{\a_{,r}}{\a \b}, W_7 = W_6 - \frac{\e_{,r} r}{\a \e} + \frac{\b_{,r} r}{\a \b}+\frac{1}{\a}. \label{SO3_NoT1} \eeq

Imposing these conditions the axial part and vector part of the torsion are then
\beq \begin{aligned} {\bf A} & = 2 W_2 \bh^1 + 2 W_1 \bh^2, \\ 
{\bf V} & = -3 \left( W_6 + \frac{\b_{,r} r}{\a \b}
+\frac{1}{\a} \right) \bh^1 - 3 \left( W_3 + \frac{\e_{,r}}{\b \e}\right) \bh^2. \end{aligned} \eeq

\noindent Applying a boost in the $\bh^1 - \bh^2$ plane, there exists a choice of the boost parameter $B =e^{f(r)}$ where these quantities are now
\beq {\bf A} = 2 C_1 {\bh'}^1, \quad {\bf V} = -3C_2 {\bh'}^1 \eeq  
\noindent for some real-valued constants $C_1$ and $C_2$ relative to the boosted frame $\{\bh_a'\}_{a=1}^4$. From which we determine the following conditions
\beq \begin{aligned} & W_1 = C_1 \sinh(f(r)), W_2 = C_1 \cosh(f(r)), \\&  W_3 = -\frac{\e_{,r}}{\b \e} + C_2 \sinh(f(r)), W_6 = - \frac{r \b_{,r}}{\a \b} - \frac{1}{\a} + C_2 \cosh(f(r)). \end{aligned} \label{SO3_AV_case1} \eeq 

In the remainder of this section, we will consider quantities relative to the boosted frame. We will determine conditions on the remaining free functions using the Ricci tensor and the covariant derivative of  \beq {\bf \tilde{V}} = {\bh^1}' = (2C_1)^{-1} {\bf A} = (-3C_2)^{-1} {\bf V}. \eeq

Computing the Ricci tensor, to satisfy the linear isotropy condition in equation \eqref{Ric3Iso}: 
\beq {\bf R} = R_1 ({\bf g} +  {\bf \tv} {\bf \tv}) + R_2 {\bf \tv} {\bf \tv}, \label{SO3:RicGeneral}\eeq
\noindent with $R_1$ and $R_2$ real valued constants relative to the boosted frame, one of the following conditions must be satisfied:

\beq f = \frac14 \ln \left( \frac{(r \b+\a)^2}{(r \b-\a)^2} \right), \text{ or } D_0 \e_{,r} = \a \b. \eeq

\noindent This gives two further sub-cases. 

\begin{enumerate}
\item $ D_0 \e_{,r} =  \a \b $:

Rescaling the arbitrary functions $\a$ and $\b$ so that $D_0 = 1$, we will consider the components of $\nabla_a \tv_b$ and impose the form in equation \eqref{DV3IsoST} so that
\beq \nabla {\bf \tv} =  C_3 ({\bf g} + {\bf \tv} {\bf \tv}) + C_4  {\bf \tv} {\bf \tv}. \label{SO3:DV} \eeq
\noindent A necessary constraint for this condition is then
\beq 2r \b \sinh(f) - 2\a \cosh(f) = D_1, \label{SO3_split_c1a} \eeq

\noindent where $D_1$ is an arbitrary constant of integration. It is natural to consider two further branches for when $f \neq 0$ and when $f = 0$. 

\begin{enumerate}
\item If $f=0$ then we can absorb the additional values and write $\a = D_1$. Examining the remaining conditions for $\nabla_a \tilde{V}_b$ to take the required form we find that $C_3 \neq 2 C_2$, $C_4 = 0$ and the functions $\b$ and $\e$ take the form:
\beq  \b && =  D_2 r^{\frac{D_1(2C_2 - C_3)-2}{2}},\label{SO3_bfcn_c1a0} \\ 
\e && = \frac{2 D_2 r^{\frac{D_1(2C_2 - C_3)}{2}}}{2C_2 - C_3}.  \label{SO3_efcn_c1a0}\eeq

\noindent Without loss of generality we can set $D_2 = 1$ by rescaling $r$. Returning to the Ricci tensor, we find an additional necessary condition $D_1 = \pm 1$. Plugging this into the Ricci tensor, we find the following expression:
\beq \begin{aligned} {\bf R} &= \left[-2C_1^2 -\frac14 C_3 (2C_2 - 3 C_3) \right] ({\bf g} + {\bf \tv} {\bf \tv}) + \frac{3}{4} C_3 (2C_2 - C_3) {\bf \tv} {\bf \tv}. \end{aligned} \label{SO3_Ric_C1a0} \eeq

\noindent while the co-Ricci tensor is of the form
\beq 
	{\bf \bar{R}} &= -2 C_1 (C_2 -  C_3) ({\bf g} + {\bf \tv} {\bf \tv}) - 3 C_1 C_3  {\bf \tv} {\bf \tv} \label{SO3_coRic_C1a0}
\eeq

\noindent Computing the Weyl tensor, we find that it vanishes automatically.

 \item If $f \neq 0$, equation \eqref{SO3_split_c1a} implies that $\a$ and $\b$ take the form: 
\beq \a = \frac{D_1 g(r)}{2 \cosh(f)}, \quad \b = \frac{D_1 (1-g(r))}{2 r \sinh(f)}. \label{SO3_ab_c1a} \eeq  

\noindent Necessary and sufficient conditions for $\nabla {\bf \tv}$ to take the form in equation \eqref{SO3:DV} are then $C_4 = 0$ and an algebraic equation for $g(r)$:
\beq g = \frac{- \e \sinh(2f) (C_3-2C_2) }{2 D_1} + \frac{1 + \cosh(2f)}{2}. \label{SO3_g_c1a}\eeq

\noindent Using this expression, we can put this into the differential equation for $\e$, $\e_{,r} = \a \b$ and determine the boost parameter algebraically:
\beq e^{2f} = \frac{8 \e_{,r} r \pm \sqrt{\e^4 (C_3 - 2C_2)^4 - 2 D_1^2 \e^2 (C_3 - 2C_2)^2 + 64 \e_{,r}^2 r^2 + D_1^4 }}{D_1^2 + \e^2 (C_3 - 2C_2)^2-4\e (C_3 - 2C_2)}. \label{SO3_f_c1a} \eeq

Using these expressions for the Ricci tensor, it takes the form in equation \eqref{SO3:RicGeneral} only if $D_1 = \pm 2 $. Imposing this condition we find:
\beq  \begin{aligned} {\bf R} =& \left[-2C_1^2 -\frac14 C_3 (2C_2 - 3 C_3)\right] ({\bf g} +  {\bf \tv} {\bf \tv}) + \frac{3}{4} C_3 (2C_2 - C_3)  {\bf \tv} {\bf \tv}, \end{aligned}  \label{SO3_Ric_c1a} \eeq

\noindent while the co-Ricci tensor is of the form
\beq 
\begin{aligned}
{\bf \bar{R}} &= -2C_1(C_2-C_3) ({\bf g} +  {\bf \tv} {\bf \tv}) - 3C_1 C_3  {\bf \tv} {\bf \tv}. \end{aligned} \label{SO3_coRic_c1a}
\eeq

\noindent Lastly, with these constraints the Weyl tensor vanishes. 

\end{enumerate}

\item $f = \frac14 \ln \left( \frac{(r \b+\a)^2}{(r \b-\a)^2}\right)$:

In this case, the linear isotropy conditions for the Ricci tensor and $\nabla {\bf \tv}$ can be solved. However, the Weyl tensor has non-vanishing components, regardless of the choice of the arbitrary functions in the coframe and boost parameter. Thus, the linear isotropy condition cannot be imposed and at most a $G_6$ is permitted.

\end{enumerate}

\subsubsection{Case 2} \label{subsec:SO3 Case 2}

The vierbein relative to the coordinates $\{ t, r, \theta, \phi\}$ is now 
\beq h^a_{~\mu} = \left[ \begin{array}{cccc} \a & 0 & 0 & 0 \\ 0 & \b & 0 & 0 \\ 0 & 0 & \e & 0 \\ 
0 & 0 & 0 & \e \end{array} \right], \label{SO3_VB_c2} \eeq

\noindent where $\e = r$ or $\e = \e_0$ a constant. With $\e = \e_0$, the Weyl tensor never vanishes, and so the analysis of this case will be excluded \cite{McNutt2024}.  Using the frame with $\e = r$, along with the connection in equation \eqref{SO3connection}, we can compute the curvature tensor and torsion tensor. As before, to ensure a 3 dimensional linear isotropy group, the tensor part of the torsion must vanish, giving the following conditions:
\beq W_4 = W_2,~W_8=W_1,~W_5 = W_3 + \frac{1}{\b r}- \frac{\a_{,r}}{\a \b},~ W_7 = W_6. \label{SO3_NT2} \eeq

\noindent The remaining axial part and vector part of the torsion are then
\beq \begin{aligned}  {\bf A} &=  2 W_2 \bh^1 + 2 W_1 \bh^2, \\ 
{\bf V} & = -3W_6 \bh^1 - 3\left(W_3 - \frac{1}{r \b} \right) \bh^2. \end{aligned} \label{SO3_AV2} \eeq

By making a boost in the $\bh_1 - \bh_2$ plane with parameter $B =e^{f(r)}$, we have a new frame basis ${\bh'}^1$ where ${\bf A} = 2C_1 {\bh'}^1$ and ${\bf V} = -3C_2 {\bh'}^1$ for some real-valued constants $C_1$ and $C_2$. We will consider the one-form \beq {\bf \tilde{V}} = {\bh^1}' = (2C_1)^{-1} {\bf A} = (-3C_2)^{-1} {\bf V}.\eeq  This then gives further conditions on the arbitrary functions:\beq \begin{aligned} & W_1 = C_1 \sinh(f), W_2 = C_1 \cosh(f), \\
& W_3 =-\frac{1}{r \b}+ C_2 \sinh(f), W_6 = C_2 \cosh(f). \end{aligned} \label{SO3_AV_c2} \eeq

\noindent Relative to the boosted frame, we will determine conditions on the remaining free functions using the covariant derivatives of ${\bf \tilde{V}} = {\bh^1}'$ and the Ricci tensor. If the Ricci tensor satisfies the condition in equation \eqref{SO3:RicGeneral}, we again find that one of the following conditions must be satisfied:
\beq f = 0 \text{ or }   \a \b = D_0. \eeq

\noindent We will consider each condition separately as sub-cases.
\begin{enumerate}

\item $\a \b = D_0$

Rescaling the arbitrary functions $\a$ and $\b$ to set $D_0 = 1$, we will now impose the linear isotropy condition on $\nabla_a \tilde{V}_b$ and $R_{ab}$ relative to the boosted frame. As in the previous case, we will consider if $f = 0 $ or not as branches. 

\begin{enumerate}
\item In the $f \neq 0$ case, requiring that $\nabla {\bf \tv}$ takes the form in equation \eqref{SO3:DV} implies that the vierbein functions are of the form: 
\beq \a = -\frac{D_1}{2\cosh(f)},~ \b =- \frac{2\cosh(f)}{D_1},~ \e = r. \label{SO3_abefcn_c2a2i} \eeq 

\noindent We also find conditions on the constants $C_3 \neq 2 C_2$ and $C_4 = 0$ and an explicit expression for the boost parameter:  
\beq f = \frac12 \ln \left( -\frac{r(2C_2 - C_3) + D_1 }{r(2C_2 - C_3) - D_1} \right). \label{SO3_ffcn_c2a2i} \eeq

The Ricci tensor takes the form in equation \eqref{SO3:RicGeneral} when $D_1 = \pm 2$. Hence, $\nabla {\bf \tv}$ and the Ricci tensor take the form:
\beq \begin{aligned} \nabla {\bf \tv} &= C_3 ({\bf g} +  {\bf \tv} {\bf \tv}), \\
{\bf R} &= \left[-2C_1^2-\frac14 C_3(2C_2 - 3C_3) \right]({\bf g} +  {\bf \tv} {\bf \tv}) \quad + \frac{3}{4} (2C_2 - C_3) C_3  {\bf \tv} {\bf \tv}, \label{SO3_DVRic_c2a2i} \end{aligned} \eeq 
\noindent while the co-Ricci tensor is of the form
\beq 
\begin{aligned}
{\bf \bar{R}} &= -2C_1(C_2-C_3)({\bf g} +  {\bf \tv} {\bf \tv}) - 3C_1C_3  {\bf \tv} {\bf \tv}. \label{SO3_coDVRic_c2a2i} \end{aligned} \eeq 

\noindent In this case the Weyl tensor now vanishes, giving no further conditions.

\item If $f = 0$ then requiring that $\nabla {\bf \tv}$ is of the form in equation \eqref{SO3:DV}, implies that the function $\a$ must be a real-valued constant, $\a = D_1$. It follows that $1 = \a \b$ requires that $\b$ must also be constant. Thus the vierbein components are: 
\beq \a = D_1,~ \b = D_1^{-1}, \text{ and } \e = r. \label{SO3_abefcn_c2a1i} \eeq 

For these vierbein functions, it follows that $\nabla {\bf \tv}$, the Ricci tensor and the co-Ricci tensor take the form in equations \eqref{SO3_DVRic_c2a2i} and \eqref{SO3_coDVRic_c2a2i} where $C_3 = 2C_2$. The Weyl tensor vanishes as well. This is in fact a subcase of the previous case.

\end{enumerate}

\item $\a \b \neq$ constant and $f = 0$ 

Imposing equation \eqref{SO3:DV} in this case, we find that $\a = D_1$, a real-valued constant, $C_3 = 2C_2$ and $C_4 = 0$. Thus the covariant derivative of ${\bf \tv}$ is: 
\beq \nabla {\bf \tv} = 2 C_2 ({\bf g} +  {\bf \tv} {\bf \tv}). \label{SO3_DV_c2bi} \eeq 

\noindent Similarly, the Ricci tensor takes the form in equation \eqref{SO3:RicGeneral} with $C_6 = 0 $  when $\b$ is of the form: 
\beq \b = \pm \frac{1}{\sqrt{D_2 r^2 +1}}, \label{SO3_bfcn_c2bi} \eeq
\noindent yielding the following Ricci tensor:
\beq {\bf R} = (-2C_1^2 +2 C_2^2 - 2 D_2) ({\bf g} +  {\bf \tv} {\bf \tv}). \label{SO3_R_c2bi} \eeq
\noindent The co-Ricci tensor takes the form:
\beq
\begin{aligned}
{\bf \bar{R}} &= (2 C_1 C_2) ({\bf g} +  {\bf \tv} {\bf \tv}) - 6 C_1 C_2  {\bf \tv} {\bf \tv}.
\end{aligned}  \label{SO3_coR_c2bi} 
\eeq
\noindent The Weyl tensor in this case is trivial in this case. The time coordinate can be rescaled to set $D_1 = \pm 1$, depending on the sign of $D_1$.

\end{enumerate}

\subsection{$SO(1,2)$ isotropy} \label{sec:SO12}

\subsubsection{Case 1}\label{subsec:SO12 Case 1}

Prior to choosing the symmetry adapted coordinates, we may rescale $z$ to set $C_0 = 1$. The vierbein relative to the coordinates $\{ t, \theta, \phi, z\}$ takes the form 
\beq h^a_{~\mu} = \left[ \begin{array}{cccc} {\a(t)} & 0 & 0 & \a(t) t \\ 
0 & \b(t) & 0 & 0 \\ 0 & 0 & \b(t) \sinh(\theta) & 0 \\ 0 & 0 & 0 & \e(t) \end{array} \right]. \label{SO12_VB_c1}  \eeq

\noindent With the connection in equation \eqref{SO12connection}, we can compute the curvature tensor and torsion tensor. Imposing the vanishing of the tensor-part of the tensor, we find the following conditions,
\beq \begin{aligned} & W_4 = -W_2,~ W_8 = W_1,~ W_5 = - W_3 - \frac{\b_{,t} t}{\b \e} + \frac{\a_{,t} t}{\a \e }+ \frac{1}{  \e}, \\ 
&W_7 = W_6 - \frac{\b_{,t}}{\a \b } + \frac{\e_{,t}}{\a \e }. \end{aligned} \label{SO12_NT1} \eeq

The axial and vector parts of the torsion are then 
\beq \begin{aligned} 
{\bf A} &= 2 W_2 \bh^1 + 2 W_1 \bh^4, \\
{\bf V} &= -3 \left( W_6 + \frac{\e_{,t}}{\a \e }  \right) \bh^1 +3 \left( W_3 + \frac{\b_{,t} t}{\b \e } \right) \bh^4
\end{aligned} \label{SO12_AV1} \eeq

To preserve the linear isotropy condition, then relative to a new frame $\{ \bh'_a\}_{a=1}^4$ produced by boosting in the $\bh_1 - \bh_4$ plane, we must have ${\bf A} = 2C_1 {\bh^4}'$ and ${\bf V} = 3C_2 {\bh^4}'$ for some real-valued constants $C_1$ and $C_2$. With the boost parameter $B = e^{f(t)}$, we find further conditions on the components of the connection:
\beq \begin{aligned} 
& W_1 = C_1 \cosh(f(t)), ~ W_2 = C_1 \sinh(f(t)), \\
& W_3 = - \frac{\b_{,t} t}{\b \e } + C_2 \cosh(f(t)),~ W_6 = - \frac{\e_{,t}}{\a \e}-C_2 \sinh(f(t)). 
\end{aligned} \label{SO12_AV_c1} \eeq
\noindent Working in the boosted frame, we can now determine conditions on the remaining free functions using the Ricci tensor and the covariant derivatives of \beq {\bf \tilde{V}} = \bh^4 = (2C_1)^{-1} {\bf A} = (3C_2)^{-1} {\bf V}. \eeq

Looking at the components of the Ricci tensor, the following component must vanish to satisfy equation \eqref{Ric3Iso}: 
\beq R_{41} = \frac{e^{2f} (e^{-4f} (\a t-\e)^2 - (\a t+\e)^2 ) (\a \e \b_{,t,t} - \a \b_{,t} \e_{,t} - \e \b_{,t} \a_{,t})}{2\a^3 \b \e^3}. \eeq

\noindent This vanishes if $f = - \frac{1}{4} \ln \left( \frac{(\a t + \e)^2}{(\a t - \e)^2} \right)$ or $D_0 \b_{,t} = \a \e$. 

\begin{enumerate}

\item $D_0 \b_{,t} = \a \e$

The arbitrary functions of $\a$ and $\b$ can be rescaled so that $D_0 = 1$. Then, imposing the linear isotropy condition on $\nabla_a \tv_b$ and $R_{ab}$ relative to the boosted frame, we find a helpful splitting condition from integrating $\nabla_1 \tv_4 = 0$ with respect to $t$: 
\beq - 2t \a \sinh(f) - 2\e \cosh(f) = D_1, \label{SO12_split_c1a} \eeq

\noindent where $D_1$ is an arbitrary constant of integration.  We will consider the cases $f = 0$ and $f \neq 0$ separately. 

\begin{enumerate}

\item If $f \neq 0$, the algebraic condition in equation \eqref{SO12_split_c1a} implies that $\a$ and $\e$ take the form: 
\beq \a = -\frac{D_1 g(t)}{2 t \sinh(f)}, \e = -\frac{D_1 (1-g(t))}{2 \cosh(f)}. \label{SO12_ae_c1_1b} \eeq  

Then, requiring that $\nabla_1 \tv_1 = - \nabla_2 \tv_2 = - \nabla_3 \tv_3  = C_3$, we find an algebraic equation for $g(t)$, 
\beq g(t) = \frac{-\b \sinh(2f) (2C_2 - C_3)}{2D_1} + \frac{1-\cosh(2f)}{2}. \label{SO12_gfcn_c1_1b} \eeq  

\noindent The differential equation, $\b_{,t} = \a \e$ allows for the boost parameter to be written algebraically: 
\beq &e^{2f} = \frac{-8 \b_{,t} t \pm \sqrt{\b^4 (C_3-2C_2)^4 - 2 D_1^2 \b^2 (C_3-2C_2)^2 + D_1^4 + 64 \b_{,t}^2 t^2}}{ \b^2 (C_3-2C_2)^2 - 2  D_1 \b (C_3-2C_2) + D_1^2}&. \label{SO12_ffcn_c1_1b} \eeq

Computing the Ricci tensor, the linear isotropy condition is satisfied if 
\beq R_{11} + R_{22} = \frac{D_1^2 - 4 }{4\b} = 0. \eeq

\noindent Hence, $D_1 = \pm 2 $, and the Ricci tensor is of the form
\beq \begin{aligned} {\bf R} =& \left( 2 C_1^2 - \frac14 C_3 (3C_3-2C_2 ) \right) ({\bf g} - {\bf \tv} {\bf \tv})- \frac34 C_3 ( 2C_2 - C_3)  {\bf \tv} {\bf \tv}, \end{aligned} \label{SO12_Ric_c1_1b} \eeq

\noindent while the co-Ricci tensor is
\beq \begin{aligned} {\bf \bar{R}} =&- 2C_1(C_2-C_3) ({\bf g} - {\bf \tv} {\bf \tv}) + 3 C_1 C_3  {\bf \tv} {\bf \tv}, \end{aligned} \label{SO12_coRic_c1_1b} \eeq

\noindent With these conditions the Weyl tensor identically vanishes.

\item If $f = 0 $ then from equation \eqref{SO12_split_c1a}, we may write $\e= D_1$. The remaining non-vanishing components of $\nabla_a \tv_b$ are 
\beq \begin{aligned} 
& \nabla_1 \tv_1 = -2 C_2 + \frac{2\a_{,t} t}{\a D_1 } + \frac{2 \a}{\a D_1 }, \\
& \nabla_2 \tv_2 = \nabla_3 \tv_3 = 2 C_2 - \frac{\a t}{  \b}. 
\end{aligned} \eeq

\noindent In the case of $SO(1,2)$ linear isotropy, these components must be constant and satisfy $-\nabla_1 \tv_1 = \nabla_2 \tv_2 = C_3$. Note that the second equation $\nabla_2 \tv_2 = C_3$ and the requirement that $\b$ is non-zero then implies that $C_3 \neq 2C_2$. 

Solving the resulting differential equation for $\a$ gives: 
\beq \begin{aligned}
& \a = D_2 t^{\frac{D_1(2C_2 - C_3)-2}{2}},
\end{aligned} \label{SO12_afcn_c1_1a}\eeq

\noindent then solving for $\b$ from the remaining equation gives
\beq \begin{aligned}
& \b = 2\frac{D_1 t^{\frac{ D_1(2 C_2 - C_3)}{2}}}{(2C_2 - C_3)}.
\end{aligned} \label{SO12_bfcn_c1_1a} \eeq

\noindent By rescaling the $t$ coordinate, we may set $D_2 = 1$.  Computing the Ricci tensor, and requiring that $R_{11} + R_{22} = 0$, we find that 
\beq D_1^2 = 1. \eeq

\noindent Using these values for the constants, the Weyl tensor now vanishes identically, giving no further conditions. Finally, the Ricci tensor is of the form:
\beq \begin{aligned} 
& {\bf R} = \left(2C_1^2 -\frac14 C_3(2C_2 - 3 C_3) \right) ({\bf g} - {\bf \tv} {\bf \tv}) + \frac34 C_3 ( 2 C_2 - C_3)  {\bf \tv} {\bf \tv},
\end{aligned} \label{SO12_R_c1_1a} \eeq
\noindent and the co-Ricci tensor is
\beq \begin{aligned} 
& {\bf \bar{R}} = -2 C_1(C_2 - C_3) ({\bf g} - {\bf \tv} {\bf \tv}) + 3 C_1 C_3  {\bf \tv} {\bf \tv}.
\end{aligned} \label{SO12_coR_c1_1a} \eeq

\end{enumerate}

\item $f = - \frac{1}{4} \ln \left( \frac{(\a t + \e)^2}{(\a t - \e)^2} \right)$

Computing $\nabla_a \tv_b$, the condition $\nabla_1 \tv_1 + \nabla_2 \tv_2 = 0$ is satisfied if
\beq \b = D_1 \sqrt{\e^2 - t^2 \a^2}. \nonumber \eeq
However, substituting this into the Ricci tensor, we see that one of the necessary conditions for $SO(1,2)$ linear isotropy is never satisfied since
\beq R_{11} + R_{22} = \frac{1}{D_1 (t^2 \a^2 - \e^2)}. \eeq

\noindent We conclude there are no $G_7$ geometries within this branch.

\end{enumerate}

\subsubsection{Case 2 }\label{subsec:SO12 Case 2}

The vierbein relative to the coordinates $\{ t, \theta, \phi, z\}$ is
\beq h^a_{~\mu} = \left[ \begin{array}{cccc} \a & 0 & 0 & 0 \\ 
0 & \b & 0 & 0 \\ 
0 & 0 & \b \sinh(\theta) & 0 \\ 
0 & 0 & 0 & \e \end{array} \right], \label{SO12_VB_c2}\eeq

\noindent where $\b =t$ or $\b = \b_0$, a constant. With the above coframe and the connection given in equation \eqref{SO12connection}, the curvature tensor and torsion tensor can be computed. Requiring that the tensor part of the torsion tensor vanishes yields
\beq W_4 = -W_2,~ W_8 = W_1,~ W_5 = -W_3,~W_7 = W_6 - \frac{1}{\a t} + \frac{\e_{,t}}{\a \e}. \label{SO12_NT2} \eeq

\noindent The axial and vector parts of the torsion tensor are now
\beq \begin{aligned} 
& {\bf A } = 2 W_2 \bh^1 + 2 W_1 \bh^4, \\
& {\bf V} = -3\left( W_6 + \frac{\e_{,t}}{\a \e} \right) \bh^1 + 3  W_3  \bh^4 . \end{aligned} \label{SO12_AV_c2} \eeq

Relative to a boosted frame $\{ \bh_a' \}_{a=1}^4$, the linear isotropy condition implies that ${\bf A} = 2C_1 {\bh^4}'$ and ${\bf V} = 3C_2 {\bh^4}'$ for some real-valued constants $C_1$ and $C_2$. We will work with \beq {\bf \tilde{V}} = {\bh^4}' = (2C_1)^{-1} {\bf A} = (3C_2)^{-1} {\bf V}. \eeq 
\noindent The boost parameter is then $B = e^{f(t)}$ and we can rewrite the remaining components of the connection as
\beq \begin{aligned}
& W_1 = C_1 \cosh(f),~W_2 = C_1 \sinh(f),\\
& W_3 =  C_2 \cosh(f),~~ W_6 = - \frac{\e_{,t}}{\a \e } - C_2 \sinh(f). \end{aligned} \eeq

\noindent Working in the boosted frame, we will determine conditions on the remaining free functions.

In the case that $\b$ is a constant, the linear isotropy condition forces $\e$ to be constant. With this condition on $\e$, the Weyl tensor will have non-vanishing components, and so no $G_7$ geometries lie within this subclass. In the case that $\b =t$, we again find an algebraic constraint for the remaining vierbein functions by looking at the components of the Ricci tensor.  

Setting $\beta = t$, the following component of the Ricci tensor must vanish to satisfy equation \eqref{Ric3Iso}: 
\beq R_{41} = \frac{e^{2f} (e^{-4f} - 1 ) (- \a \e_{,t} - \e \a_{,t})}{\a^3 t \e}. \eeq

\noindent This vanishes if $f = 0$ or $ D_0 \b_{,t} = \a \e $. 

\begin{enumerate}

\item $\a \e = D_0$

We will rescale the arbitrary functions to set $D_0 = 1$. Taking $\nabla_1 \tv_4 = 0$ and integrating with respect to $t$, we have the following condition: 
\beq - 2\cosh(f) \e = D_1. \label{SO12_split_c2a} \eeq

Using equation \eqref{SO12_split_c2a} along with the algebraic equation $\a \e = 1$ we find the following
\beq \a = - \frac{\cosh(f)}{D_1},~~\e = - \frac{D_1}{\cosh(f)}. \label{SO12_ae_c21a } \eeq

\noindent If we require $\nabla_1 \tv_1 + \nabla_2 \tv_2 = 0$, then this determines the boost parameter:
\beq e^{2f} = -\frac{ D_2 t +1}{D_2 t-1}. \label{SO12_f_c21a} \eeq

\noindent Using this, we find that 
\beq \nabla {\bf \tv} = (-2D_1 D_2-2C_2 ) ( {\bf g} - {\bf \tv} {\bf \tv}). \label{SO12_DV_c21a} \eeq

 Similarly, imposing $R_{11} + R_{22} = 1$ gives
\beq D_1 = \pm 1, \eeq
\noindent and the Ricci tensor takes the form 
\beq \begin{aligned} {\bf R} =& (2C_1^2 - 2 C_2^2 -  5 C_2 D_1 D_2 -3 D_2^2) ( {\bf g} - {\bf \tv} {\bf \tv}) - 3 D_2(C_2 + D_2 D_1)  {\bf \tv} {\bf \tv} \end{aligned} \label{SO12_R_c21a} \eeq

\noindent and the co-Ricci tensor is 
\beq \begin{aligned} {\bf \bar{R}} =& 2C_1(C_2+2D_2) ( {\bf g} - {\bf \tv} {\bf \tv}) + 6 C_1(C_2+D_2)  {\bf \tv} {\bf \tv}. \end{aligned} \label{SO12_coR_c21a} \eeq
\noindent The Weyl tensor vanishes, giving no further conditions. 

\item $f = 0$ and $\a \e \neq$ constant

In this case, the vanishing of $\nabla_1 \tv_4$ gives the differential constraint 
\beq 2\frac{\e_{,t}}{\a \e}=0 \eeq
\noindent which implies that $\e = D_1$ is a constant, and then by rescaling $z$ we may set $D_1 = \pm1$ depending on the sign of $D_1$. The remaining components satisfy $-\nabla_1 \tv_1 = \nabla_2 \tv_2 = \nabla_3 \tv_3 = 2C_2$.  Computing the Weyl tensor the components yield the following condition on $\a$
\beq -\a^3+ \a_{,t}+\a = 0. \eeq

\noindent We may solve for the vierbein functions completely
\beq \a = \pm \frac{1}{\sqrt{1+D_2 t^2}}, \b = t, \e = 1. \label{SO12_ae_c2b} \eeq

\noindent Using these expressions, the Ricci tensor is then
\beq {\bf R} = (2C_1^2 -2C_2^2 + 2D_2)({\bf g} - {\bf \tv} {\bf \tv}), \label{SO12_R_c2b}  \eeq
\noindent while the co-Ricci tensor is
\beq {\bf \bar{R}} = 2 C_1 C_2({\bf g} - {\bf \tv} {\bf \tv}) + 6C_1 C_2 {\bf \tv} {\bf \tv}. \label{SO12_coR_c2b}  \eeq
\end{enumerate}

\subsection{$E(2)$ isotropy} \label{sec:E2}

Following the remark in section \ref{sec:G4spaces}, the vierbein relative to the coordinates $\{ u, v, r, \theta\}$ take the form:
\beq k^a_{~\mu} = \left[ \begin{array}{cccc} 1 & 0 & 0 & 0 \\ 
1 & \b(u) & 0 & 0 \\ 0 & 0 & \e(u) & 0 \\ 0 & 0 & 0 & \e(u) r \end{array} \right]. \label{E2_VB}  \eeq

\noindent Using this along with the connection in equation \eqref{E2connection} we can compute the curvature tensor and torsion tensor. Following from the discussion in section \ref{sec:CKalg}, we ask that the tensor-part of the torsion vanishes, which gives the following conditions:
\beq W_5 = W_2,~W_7 = W_1,~ W_8 = W_6,~ W_4 = -W_3 - \frac{\e_{,u}}{\e} + \frac{\b_{,u}}{\b}, \label{E2_NT} \eeq

\noindent Imposing these conditions, the axial and vector parts of the torsion are
\beq \begin{aligned} 
& {\bf A} = -2 W_1 \bk^1 +2 W_2 \bk^2,\\
& {\bf V} = 3 \left(W_3 - \frac{\b_{,u}}{\b} \right) \bk^1 -3 W_6 \bk^2, 
\end{aligned} \eeq 

\noindent In this case, boosts will not affect the form of these one-forms significantly, while null rotations will further complicate the expressions. Imposing the linear isotropy condition: $|{\bf A}|^2 = |{\bf V}|^2 = 0$ and ${\bf A}$ and ${\bf V}$ proportional to $\bk_1$, yields two more conditions
\beq W_1 = C_1,~W_2 = 0,W_3 = \frac{\b_{,u}}{\b}+C_2~~W_6 = 0. \label{E2_AV} \eeq

We will again consider $\tilde{V} = \bk^1 = (-2C_1)^{-1} {\bf A} = (3 C_2)^{-1} {\bf V}$ and compute $\nabla_a \tv_b$.   There is one non-zero component that must be constant, $\nabla_1 \tv_1 = C_3$ giving the differential equation
\beq -2C_2 -\frac{2\b_{,u}}{\b} = C_3, \eeq

\noindent which has the solution
\beq \b = D_1 e^{\frac{(2C_2 +C_3)u}{2}}. \label{E2_b} \eeq

Computing the Ricci tensor, there is only one constant component, \beq {\bf R} = C_4 \bk^1 \bk^1, \label{E2_Ric} \eeq

\noindent which gives the following differential constraint for $\e$:
\beq \frac{\e_{,u} (2 C_2 + C_3) + 2 \e_{,u,u}}{\e} = C_4, \eeq

\noindent This equation has the general solution
\beq \e = D_2 e^{\left(-\frac14 (2C_2 +C_3) + \sqrt{(2C_2 + C_3)^2 - 8 C_4} \right) u} + D_3 e^{\left(-\frac14 (2C_2 +C_3) - \sqrt{(2C_2 + C_3)^2 - 8 C_4} \right) u}.  \label{E2_e} \eeq

\noindent Imposing this condition, the co-Ricci tensor is then 
\beq {\bf \bar{R}} = C_1 (2C_2 - C_3) \bk^1 \bk^1  \label{E2_coRic}. \eeq

\noindent With these conditions the Weyl tensor vanishes automatically. 

There is another possible case within this branch. If we instead suppose that $|{\bf A}|^2 = |{\bf V}|^2 = 0$ but require that ${\bf A}$ and ${\bf V}$ are instead proportional to $\bk_2$ then we have the following conditions
\beq W_1 = 0, W_2 = C_1, W_3 = \frac{\b_{,u}}{\b}, W_6 = C_2. \label{E2_AV_alt} \eeq
\noindent Then taking ${\bf \tilde{V}} = \bk^2 = (2C_1)^{-1} {\bf A} = (3C_2)^{-1} {\bf V}$, the isotropy condition for $\nabla_a \tilde{V}_b$ requires that
\beq \frac{\b_{,u}}{\b} =0, ~\frac{\e_{,u}}{\e} =0, C_2 = 0. \eeq
\noindent It then follows that $\nabla {\bf \tilde{V}} = 0$ and the curvature tensor vanishes entirely. This describes a Minkowski spacetime with non-vanishing axial torsion, arising from a non-trivial spin-connection. 

\section{Summary} \label{sec:summary}

With the analysis concluded, we summarize the defining quantities for each of the $G_7$ geometries for each linear isotropy group and sub-divided by the possible forms for the fourth affine frame symmetry generator in the original coordinate system where the vierbein matrix is diagonal. 

While the previous analysis concentrated on the geometrical development of Riemann-Cartan geometries, it is of some interest to also consider the physical aspects of such geometries in gravitational models.  In order to make conclusions about the physical aspects of these geometries, one is required to assume a gravitational theory.  For example in GR and Einstein-Cartan theory, the Ricci tensor is sufficient to describe the energy-momentum tensor, which is proportional to the Einstein tensor [28, 29]. Here, as a simple illustration, we shall assume an Einstein-Cartan theory of gravity in which the action is $$S=\int d^4x \left( \sqrt{|g|} \frac{R}{2\kappa} +L_{m}  \right)$$ where $R$ is the Ricci scalar, and $L_{m}$ is the Lagrangian density for matter.  We note that the torsion in such Einstein-Cartan theories only couples to the spin of the matter and does not propagate.  Other theories of gravity in which the Lagrangian for the gravity field depends on functions of the Ricci scalar, or functions of the scalar invariants of the torsion, or both, will have different conclusions concerning the nature of the energy momentum tensor.  However, some general conclusions about the energy-momentum tensor can be drawn from
the form of the Ricci tensor for a generic Riemann-Cartan geometry with non-vanishing
curvature tensor and torsion tensor assuming an Einstein-Cartan theory of gravity.  
\vspace{ 3 mm }

\noindent $SO(3)$: Case 1, ${\bf Y} = \partial_t + r \partial_r$ with \beq |{\bf Y}|^2 = - \a(r e^{-t})^2 + r^2 \b(r e^{-t})^2 <0. \eeq

\noindent Then relative to adapted coordinates, the vierbein is of the form \eqref{SO3_VB_c1} and the connection is given by \eqref{SO3connection} with the conditions \eqref{SO3_NoT1} and \eqref{SO3_AV_case1}. 

When expressed in the boosted frame, the axial and vector parts of the torsion are ${\bf A} = 2 C_1 {\bf \tilde{V}}$ and ${\bf V} = -3C_2 {\bf \tilde{V}}$ where ${\bf \tilde{V}} = {\bh}_1'$. Similarly the covariant derivative of ${\bf \tilde{V}}$ can be summarized by a single scalar $C_3$. The following table displays for each case: the relevant equations for the  vierbein functions $\a, \b, \e$ and the boost parameter $f$; the values of the three scalars $C_1, C_2$ and $C_3$; the form of the Ricci tensor and co-Ricci tensor; and the permitted values of a constant of integration.

\begin{table}[h]
\centering
\beq \begin{array}{c|c|c|c|c|c|c|c|c|c}
\a & \b & \e & f & {\bf A} & {\bf V} & \nabla {\bf \tv} & {\bf R} & {\bf \bar{R}} &  D_1 \\ \hline
D_1 & \eqref{SO3_bfcn_c1a0} & \eqref{SO3_efcn_c1a0} & 0  &  0 & C_2 & C_3 \neq 2C_2 & \eqref{SO3_Ric_C1a0} & \eqref{SO3_coRic_C1a0} & \pm 1 \\
\eqref{SO3_ab_c1a} \& \eqref{SO3_g_c1a} & \eqref{SO3_ab_c1a} \& \eqref{SO3_g_c1a} & \e(r) & \eqref{SO3_f_c1a}  & C_1 & C_2 & C_3 & \eqref{SO3_Ric_c1a} & \eqref{SO3_coRic_c1a} &   \pm 2 
\end{array} \nonumber \eeq
\caption{The geometric inputs for spherically symmetric stationary solutions with the largest affine frame symmetry groups, along with their associated torsion and curvature parts}
\label{Table:SO3 Stationary}
\end{table}
These geometries are stationary spherically symmetric geometries whose source is either an isotropic perfect fluid, a cosmological constant, or a vacuum depending on the values of the constants $C_1, C_2$ and $C_3$. However, in some branches choosing the constants to impose a vanishing Ricci tensor leads to the $G_7$ Riemann-Cartan geometry simplifying to the Minkowski geometry where both the curvature tensor and the torsion tensor vanish. We note that if $C_3 = 0$, the teleparallel analogue of the de Sitter solution \cite{Coley:2022} arises as a special case of branch 1.1.a.
\vspace{2 mm}

\noindent $SO(3)$: Case 2, ${\bf Y} = \partial_t$ with \beq |{\bf Y}|^2 = - \a(r)^2  <0. \eeq

\noindent Then relative to adapted coordinates, the vierbein is of the form \eqref{SO3_VB_c2} and the connection is given by \eqref{SO3connection} with the conditions \eqref{SO3_NT2} and \eqref{SO3_AV_c2}.

When expressed in the boosted frame, the axial and vector parts of the torsion are ${\bf A} = 2 C_1 {\bf \tilde{V}}$ and ${\bf V} = -3C_2 {\bf \tilde{V}}$ where ${\bf \tilde{V}} = {\bh}_1'$. Similarly the covariant derivative of ${\bf \tilde{V}}$ can be summarized by a single scalar $C_3$. The following table displays for each case: the relevant equations for the  vierbein functions $\a, \b, \e$ and the boost parameter $f$; the values of the three scalars $C_1, C_2$ and $C_3$; the form of the Ricci tensor and co-Ricci tensor; and the permitted values of a constant of integration.
\begin{table}[h]
\centering
\beq \begin{array}{c|c|c|c|c|c|c|c|c|c}
 \a & \b & \e & f & {\bf A} & {\bf V} & \nabla {\bf \tv} & {\bf R} & {\bf \bar{R}} & D_1 \\ \hline
\eqref{SO3_abefcn_c2a2i} & \eqref{SO3_abefcn_c2a2i} & r & \eqref{SO3_ffcn_c2a2i} & 0 & C_2 & C_3 & \eqref{SO3_DVRic_c2a2i}-II & \eqref{SO3_coDVRic_c2a2i} & \pm {2} \\
 D_1 & \eqref{SO3_bfcn_c2bi} & r & 0 & C_1 & C_2 & 2 C_2 & \eqref{SO3_R_c2bi} & \eqref{SO3_coR_c2bi} &   \pm 1 \\
\end{array} \nonumber \eeq
\caption{The geometric inputs for spherically symmetric static solutions with the largest affine frame symmetry groups, along with their associated torsion and curvature parts}
\label{Table:SO3 Static}
\end{table}

The geometries in this case describe static spherically symmetric geometries. Depending on the choice of the constants $C_1, C_2$ and $C_3$, the corresponding geometries' sources are of the form of an isotropic perfect fluid, a cosmological constant, or vacuum. Geometries in branch 2.1.a can only admit a perfect fluid, otherwise they reduce to Minkowski space where the curvature tensor and torsion tensor vanish. 
\vspace{2 mm }

\noindent $SO(1,2)$: Case 1, ${\bf Y} = t \partial_t + \partial_z$ with \beq |{\bf Y}|^2 = t^2 \a(t e^{-z})^2 + \b(t e^{-z})^2 >0. \eeq

\noindent Then relative to adapted coordinates, the vierbein is of the form \eqref{SO12_VB_c1} and the connection is given by \eqref{SO12connection} with the conditions \eqref{SO12_NT1} and \eqref{SO12_AV_c1}. 

When expressed in the boosted frame, the axial and vector parts of the torsion are ${\bf A} = 2 C_1 {\bf \tilde{V}}$ and ${\bf V} = 3C_2 {\bf \tilde{V}}$ where ${\bf \tilde{V}} = {\bh}_4'$. Similarly the covariant derivative of ${\bf \tilde{V}}$ can be summarized by a single scalar $C_3$. The following table displays for each case: the relevant equations for the  vierbein functions $\a, \b, \e$ and the boost parameter $f$; the values of the three scalars $C_1, C_2$ and $C_3$; the form of the Ricci tensor and co-Ricci tensor; and the permitted values of a constant of integration.  

\begin{table}[h]
\centering
\beq \begin{array}{c|c|c|c|c|c|c|c|c|c}
\a & \b & \e & f & {\bf A} & {\bf V} & \nabla {\bf \tv} & {\bf R} & {\bf \bar{R}} & D_1 \\ \hline
\eqref{SO12_ae_c1_1b} \& \eqref{SO12_gfcn_c1_1b} & \b(t) &  \eqref{SO12_ae_c1_1b} \& \eqref{SO12_gfcn_c1_1b} &   \eqref{SO12_ffcn_c1_1b} & C_1 & C_2 & C_3 & \eqref{SO12_Ric_c1_1b} & \eqref{SO12_coRic_c1_1b}  & \pm 2 \\
\eqref{SO12_afcn_c1_1a} & \eqref{SO12_bfcn_c1_1a} & D_1 & 0 & 0 & C_2 & C_3 \neq 2C_2 & \eqref{SO12_R_c1_1a} & \eqref{SO12_coR_c1_1a} & \pm 1   \end{array} \nonumber \eeq
\caption{The geometric inputs for the $SO(1,2)$ isotropic solutions with the largest affine frame symmetry groups containing ${\bf Y} = t \partial_t + \partial_z$, along with their associated torsion and curvature parts}
\label{Table:SO12 NonDiag}
\end{table}

These geometries are necessarily stationary (whether these geometries admit static subclasses will be investigated in future work). Some of these solutions can describe geometries with vacuum or cosmological constant, except for the last case in branch 1.1.b where the vacuum condition requires that the geometry is Minkowski as the torsion tensor and curvature tensor will both vanish. Due to the $SO(1,2)$ isotropy, the non-vacuum source for such geometries can be treated as a cosmological constant acting on a three-dimensional timelike submanifold along with a scalar pressure acting on the remaining spatial direction or a cosmological constant acting on the entire space. We note that the teleparallel equivalent of anti-de Sitter space arises in branch 1.1.a when $C_3 =0$. 

\vspace{2 mm}

\noindent $SO(1,2)$: Case 2, ${\bf Y} = \partial_z$ with \beq |{\bf Y}|^2 = \b(t)^2 >0. \eeq

\noindent Then relative to adapted coordinates, the vierbein is of the form \eqref{SO12_VB_c2} and the connection is given by \eqref{SO12connection} with the conditions \eqref{SO12_NT2} and \eqref{SO12_AV_c2}. 

When expressed in the boosted frame, the axial and vector parts of the torsion are ${\bf A} = 2 C_1 {\bf \tilde{V}}$ and ${\bf V} = 3C_2 {\bf \tilde{V}}$ where ${\bf \tilde{V}} = {\bh}_4'$. Similarly the covariant derivative of ${\bf \tilde{V}}$ can be summarized by a single scalar $C_3$. The following table displays for each case: the relevant equations for the  vierbein functions $\a, \b, \e$ and the boost parameter $f$; the values of the three scalars $C_1, C_2$ and $C_3$; the form of the Ricci tensor and co-Ricci tensor; and the permitted values of a constant of integration.  
\begin{table}[h]
\centering
\beq \begin{array}{c|c|c|c|c|c|c|c|c|c}
 \a & \b & \e & f & {\bf A} & {\bf V} & \nabla {\bf \tv} & {\bf R} & {\bf \bar{R}}  &  D_1 \\ \hline
 \eqref{SO12_ae_c21a } & t & \eqref{SO12_ae_c21a } & \eqref{SO12_f_c21a} & C_1 & C_2 & C_3 = - 2D_1 D_2 - 2C_2 & \eqref{SO12_R_c21a} & \eqref{SO12_coR_c21a}  &  \pm 1 \\
 \eqref{SO12_ae_c2b} & t & D_1 & 0 & C_1 & C_2 & 2C_2 & \eqref{SO12_R_c2b} & \eqref{SO12_coR_c2b}  & \pm 1 
\end{array} \nonumber \eeq
\caption{The geometric inputs for the $SO(1,2)$ isotropic solutions with the largest affine frame symmetry groups containing ${\bf Y} = \partial_z$, along with their associated torsion and curvature parts}
\label{Table:SO12 Diag}
\end{table}
Further investigation is necessary to determine if the stationary geometries in this sub-case are static. For a non-vacuum source, the energy-momentum tensor describes a cosmological constant acting on a 3 dimensional timelike submanifold along with a pressure term on the remaining spatial direction or a cosmological constant on the entire space. Of course, vacuum solutions without a cosmological constant can be achieved for particular choices of $C_1, C_2$ and $C_3$.

\vspace{2 mm}

\noindent $E(2)$: ${\bf Y} = \partial_v$ with \beq |{\bf Y}|^2 = 0. \eeq

\noindent The vierbein is of the form \eqref{E2_VB} and the connection is given by \eqref{E2connection} with the conditions \eqref{E2_NT} and \eqref{E2_AV}. 

In this case the axial and vector parts of the torsion are ${\bf A} = -2 C_1 {\bf \tilde{V}}$ and ${\bf V} = 3C_2 {\bf \tilde{V}}$ where ${\bf \tilde{V}} = \bk_1$ or $-\bk_2$. Similarly the covariant derivative of ${\bf \tilde{V}}$ can be summarized by a single scalar $C_3$. The following table displays for each case: the relevant equations for the  vierbein functions $\a, \b, \e$ and the boost parameter $f$; the values of the three scalars $C_1, C_2$ and $C_3$; the form of the Ricci tensor and co-Ricci tensor; and the choice of ${\bf \tilde{V}}$.  

\begin{table}[h]
\centering  
\beq \begin{array}{c|c|c|c|c|c|c|c}
  \b & \e  & {\bf A} & {\bf V} & \nabla {\bf \tv} & {\bf R} & {\bf \bar{R}} & {\bf \tilde{V}}    \\ \hline
 \eqref{E2_b} & \eqref{E2_e} & C_1 & C_2 & C_3 \neq 2C_2 & \eqref{E2_Ric} & \eqref{E2_coRic} & \bk_1 \\ 
 D_1 & D_2 & C_1 & 0 & 0 & 0 & 0 & \bk_2  \end{array} \nonumber \eeq
 \caption{The geometric inputs for the $E(2)$ isotropic solutions with the largest affine frame symmetry groups, along with their associated torsion and curvature parts}
\label{Table:E2}
\end{table}

The geometries in the first row describe Riemann-Cartan geometries sourced by null radiation or are vacuum solutions without cosmological constant. We note that these geometries correspond to the $G_7$ pp-wave solutions \cite[Chapter 12, section 6]{kramer} when $C_1 =C_2 = C_3 = 0$; i.e., when the torsion tensor vanishes. The geometries in the second row describe Minkowski spacetime with constant axial torsion.

\section{Discussion} \label{sec:Discussion}

In this paper we have determined all possible Riemann-Cartan geometries which admit a 7-dimensional group of affine frame symmetries, denoted as $G_7$. To accomplish this we employed the symmetry frame formalism to determine all maximally isotropic Riemann-Cartan geometries, and determined the subclass of these geometries which admit a fourth affine frame symmetry and hence a $G_4$ group of affine frame symmetries. Then by exploiting the Cartan-Karlhede algorithm we determined conditions for the $G_4$ geometries to admit a $G_7$ and solved the resulting governing differential equations. The resulting geometries are necessarily locally homogeneous and admit the largest isotropy group permitted in Riemann-Cartan geometries having non-trivial torsion, and the potential matter sources corresponding to solutions in Einstein-Cartan theory of gravity are briefly discussed in section \ref{sec:summary}.

The $G_7$ Riemann-Cartan geometries found in this paper contain known solutions, such as the de Sitter geometry and Einstein static geometries \cite{McNutt:2023nxmnxm, Coley:2022} which are found in branch 1.1.a of the $SO(3)$ analysis where $C_3 \neq 2C_2$ and $C_3 = 0$. However, new solutions have been found such as the Riemann-Cartan analogue of anti-de Sitter space in branch  1.1.a with $C_3 \neq 2 C_2$ and $C_3 = 0$ of the $SO(1,2)$ analysis. More generally, the matter field for many of these solutions are perfect fluid solutions for $SO(3)$ isotropy and are null electromagnetic solutions for $E(2)$ isotropy. In the case of $SO(1,2)$ isotropy, less is known due to the energy-momentum tensor resembling a cosmological constant acting on a 3 dimensional timelike submanifold and a pressure term acting on the remaining spacelike direction. 

In future work, we will classify the Lie algebra structure of each of these solutions to determine whether the possible 7-dimensional Lie subgroups are realized as affine frame symmetries in Riemann-Cartan geometry and study their realizations in $f(T)$ teleparallel gravity theories together with their associated energy-momentum tensors. In addition, we note that the lower-dimensional affine frame symmetry groups are more difficult to find and require new tools to determine them using Lie algebra methods \cite{kruglikov2016}. In terms of applying the CK algorithm, the Riemann-Cartan geometries admitting a $G_5$ or $G_6$ are no longer necessarily locally homogeneous or maximally isotropic and this will complicate the analysis of the resulting differential equations. However, in the case of $G_6$ it is possible to determine geometries which admit an isotropy group that is either 2 or 3 dimensional and this will be investigated further.

\section*{Acknowledgments}
AAC and RvdH are supported by the Natural Sciences and Engineering Research Council of Canada. RvdH is supported by the Dr. W. F. James Chair of Studies in the Pure and Applied Sciences at St. Francis Xavier University. DDM is supported by the Norwegian Financial Mechanism 2014-2021 (project registration number 2019/34/H/ST1/00636).

%-------------------------------------------------------------------------
%%----------------------   SECTION       ----------------------------------
%%-------------------------------------------------------------------------

%\section*{References}
\bibliographystyle{apsrev4-2}
\bibliography{Tele-Parallel-Reference-file0}

\begin{thebibliography}{31}%
\makeatletter
\providecommand \@ifxundefined [1]{%
 \@ifx{#1\undefined}
}%
\providecommand \@ifnum [1]{%
 \ifnum #1\expandafter \@firstoftwo
 \else \expandafter \@secondoftwo
 \fi
}%
\providecommand \@ifx [1]{%
 \ifx #1\expandafter \@firstoftwo
 \else \expandafter \@secondoftwo
 \fi
}%
\providecommand \natexlab [1]{#1}%
\providecommand \enquote  [1]{``#1''}%
\providecommand \bibnamefont  [1]{#1}%
\providecommand \bibfnamefont [1]{#1}%
\providecommand \citenamefont [1]{#1}%
\providecommand \href@noop [0]{\@secondoftwo}%
\providecommand \href [0]{\begingroup \@sanitize@url \@href}%
\providecommand \@href[1]{\@@startlink{#1}\@@href}%
\providecommand \@@href[1]{\endgroup#1\@@endlink}%
\providecommand \@sanitize@url [0]{\catcode `\\12\catcode `\$12\catcode
  `\&12\catcode `\#12\catcode `\^12\catcode `\_12\catcode `\%12\relax}%
\providecommand \@@startlink[1]{}%
\providecommand \@@endlink[0]{}%
\providecommand \url  [0]{\begingroup\@sanitize@url \@url }%
\providecommand \@url [1]{\endgroup\@href {#1}{\urlprefix }}%
\providecommand \urlprefix  [0]{URL }%
\providecommand \Eprint [0]{\href }%
\providecommand \doibase [0]{https://doi.org/}%
\providecommand \selectlanguage [0]{\@gobble}%
\providecommand \bibinfo  [0]{\@secondoftwo}%
\providecommand \bibfield  [0]{\@secondoftwo}%
\providecommand \translation [1]{[#1]}%
\providecommand \BibitemOpen [0]{}%
\providecommand \bibitemStop [0]{}%
\providecommand \bibitemNoStop [0]{.\EOS\space}%
\providecommand \EOS [0]{\spacefactor3000\relax}%
\providecommand \BibitemShut  [1]{\csname bibitem#1\endcsname}%
\let\auto@bib@innerbib\@empty
%</preamble>
\bibitem [{\citenamefont {Bilenky}\ and\ \citenamefont
  {Ho{\v{s}}ek}(1982)}]{bilenky1982glashow}%
  \BibitemOpen
  \bibfield  {author} {\bibinfo {author} {\bibfnamefont {S.~M.}\ \bibnamefont
  {Bilenky}}\ and\ \bibinfo {author} {\bibfnamefont {J.}~\bibnamefont
  {Ho{\v{s}}ek}},\ }\href@noop {} {\bibfield  {journal} {\bibinfo  {journal}
  {Physics Reports}\ }\textbf {\bibinfo {volume} {90}},\ \bibinfo {pages} {73}
  (\bibinfo {year} {1982})}\BibitemShut {NoStop}%
\bibitem [{\citenamefont {Stephani}\ \emph {et~al.}(2009)\citenamefont
  {Stephani}, \citenamefont {Kramer}, \citenamefont {MacCallum}, \citenamefont
  {Hoenselaers},\ and\ \citenamefont {Herlt}}]{kramer}%
  \BibitemOpen
  \bibfield  {author} {\bibinfo {author} {\bibfnamefont {H.}~\bibnamefont
  {Stephani}}, \bibinfo {author} {\bibfnamefont {D.}~\bibnamefont {Kramer}},
  \bibinfo {author} {\bibfnamefont {M.}~\bibnamefont {MacCallum}}, \bibinfo
  {author} {\bibfnamefont {C.}~\bibnamefont {Hoenselaers}},\ and\ \bibinfo
  {author} {\bibfnamefont {E.}~\bibnamefont {Herlt}},\ }\href@noop {} {\emph
  {\bibinfo {title} {Exact Solutions of Einstein's Field Equations}}}\
  (\bibinfo  {publisher} {Cambridge University Press},\ \bibinfo {year}
  {2009})\BibitemShut {NoStop}%
\bibitem [{\citenamefont {Will}(2014)}]{Will:2014kxa}%
  \BibitemOpen
  \bibfield  {author} {\bibinfo {author} {\bibfnamefont {C.~M.}\ \bibnamefont
  {Will}},\ }\href {https://doi.org/10.12942/lrr-2014-4} {\bibfield  {journal}
  {\bibinfo  {journal} {Living Rev. Rel.}\ }\textbf {\bibinfo {volume} {17}},\
  \bibinfo {pages} {4} (\bibinfo {year} {2014})},\ \Eprint
  {https://arxiv.org/abs/1403.7377} {arXiv:1403.7377 [gr-qc]} \BibitemShut
  {NoStop}%
%%CITATION = ARXIV:1403.7377;%%
\bibitem [{\citenamefont {Weisberg}\ and\ \citenamefont
  {Huang}(2016)}]{Weisberg:2016jye}%
  \BibitemOpen
  \bibfield  {author} {\bibinfo {author} {\bibfnamefont {J.~M.}\ \bibnamefont
  {Weisberg}}\ and\ \bibinfo {author} {\bibfnamefont {Y.}~\bibnamefont
  {Huang}},\ }\href {https://doi.org/10.3847/0004-637X/829/1/55} {\bibfield
  {journal} {\bibinfo  {journal} {Astrophys. J.}\ }\textbf {\bibinfo {volume}
  {829}},\ \bibinfo {pages} {55} (\bibinfo {year} {2016})},\ \Eprint
  {https://arxiv.org/abs/1606.02744} {arXiv:1606.02744 [astro-ph.HE]}
  \BibitemShut {NoStop}%
\bibitem [{\citenamefont {{Abbott, B. P. and
  others}}(2016)}]{LIGOScientific:2016aoc}%
  \BibitemOpen
  \bibfield  {author} {\bibinfo {author} {\bibnamefont {{Abbott, B. P. and
  others}}} (\bibinfo {collaboration} {LIGO Scientific, Virgo}),\ }\href
  {https://doi.org/10.1103/PhysRevLett.116.061102} {\bibfield  {journal}
  {\bibinfo  {journal} {Phys. Rev. Lett.}\ }\textbf {\bibinfo {volume} {116}},\
  \bibinfo {pages} {061102} (\bibinfo {year} {2016})},\ \Eprint
  {https://arxiv.org/abs/1602.03837} {arXiv:1602.03837 [gr-qc]} \BibitemShut
  {NoStop}%
\bibitem [{\citenamefont {Abbott}\ \emph {et~al.}(2019)\citenamefont {Abbott}
  \emph {et~al.}}]{LIGOScientific:2018mvr}%
  \BibitemOpen
  \bibfield  {author} {\bibinfo {author} {\bibfnamefont {B.~P.}\ \bibnamefont
  {Abbott}} \emph {et~al.} (\bibinfo {collaboration} {LIGO Scientific,
  Virgo}),\ }\href {https://doi.org/10.1103/PhysRevX.9.031040} {\bibfield
  {journal} {\bibinfo  {journal} {Phys. Rev. X}\ }\textbf {\bibinfo {volume}
  {9}},\ \bibinfo {pages} {031040} (\bibinfo {year} {2019})},\ \Eprint
  {https://arxiv.org/abs/1811.12907} {arXiv:1811.12907 [astro-ph.HE]}
  \BibitemShut {NoStop}%
\bibitem [{\citenamefont {Aghanim}\ \emph {et~al.}(2020)\citenamefont {Aghanim}
  \emph {et~al.}}]{Planck:2018vyg}%
  \BibitemOpen
  \bibfield  {author} {\bibinfo {author} {\bibfnamefont {N.}~\bibnamefont
  {Aghanim}} \emph {et~al.} (\bibinfo {collaboration} {Planck}),\ }\href
  {https://doi.org/10.1051/0004-6361/201833910} {\bibfield  {journal} {\bibinfo
   {journal} {Astron. Astrophys.}\ }\textbf {\bibinfo {volume} {641}},\
  \bibinfo {pages} {A6} (\bibinfo {year} {2020})},\ \bibinfo {note} {[Erratum:
  Astron.Astrophys. 652, C4 (2021)]},\ \Eprint
  {https://arxiv.org/abs/1807.06209} {arXiv:1807.06209 [astro-ph.CO]}
  \BibitemShut {NoStop}%
\bibitem [{\citenamefont {Hinshaw}\ \emph {et~al.}(2013)\citenamefont {Hinshaw}
  \emph {et~al.}}]{WMAP:2012nax}%
  \BibitemOpen
  \bibfield  {author} {\bibinfo {author} {\bibfnamefont {G.}~\bibnamefont
  {Hinshaw}} \emph {et~al.} (\bibinfo {collaboration} {WMAP}),\ }\href
  {https://doi.org/10.1088/0067-0049/208/2/19} {\bibfield  {journal} {\bibinfo
  {journal} {Astrophys. J. Suppl.}\ }\textbf {\bibinfo {volume} {208}},\
  \bibinfo {pages} {19} (\bibinfo {year} {2013})},\ \Eprint
  {https://arxiv.org/abs/1212.5226} {arXiv:1212.5226 [astro-ph.CO]}
  \BibitemShut {NoStop}%
\bibitem [{\citenamefont {Buchert}\ \emph {et~al.}(2016)\citenamefont
  {Buchert}, \citenamefont {Coley}, \citenamefont {Kleinert}, \citenamefont
  {Roukema},\ and\ \citenamefont {Wiltshire}}]{Buchert}%
  \BibitemOpen
  \bibfield  {author} {\bibinfo {author} {\bibfnamefont {T.}~\bibnamefont
  {Buchert}}, \bibinfo {author} {\bibfnamefont {A.~A.}\ \bibnamefont {Coley}},
  \bibinfo {author} {\bibfnamefont {H.}~\bibnamefont {Kleinert}}, \bibinfo
  {author} {\bibfnamefont {B.~F.}\ \bibnamefont {Roukema}},\ and\ \bibinfo
  {author} {\bibfnamefont {D.~L.}\ \bibnamefont {Wiltshire}},\ }\href@noop {}
  {\bibfield  {journal} {\bibinfo  {journal} {International Journal of Modern
  Physics D}\ }\textbf {\bibinfo {volume} {25}},\ \bibinfo {pages} {1630007}
  (\bibinfo {year} {2016})}\BibitemShut {NoStop}%
\bibitem [{\citenamefont {Riess}\ \emph {et~al.}(2016)\citenamefont {Riess}
  \emph {et~al.}}]{Riess:2016jrr}%
  \BibitemOpen
  \bibfield  {author} {\bibinfo {author} {\bibfnamefont {A.~G.}\ \bibnamefont
  {Riess}} \emph {et~al.},\ }\href {https://doi.org/10.3847/0004-637X/826/1/56}
  {\bibfield  {journal} {\bibinfo  {journal} {Astrophys. J.}\ }\textbf
  {\bibinfo {volume} {826}},\ \bibinfo {pages} {56} (\bibinfo {year} {2016})},\
  \Eprint {https://arxiv.org/abs/1604.01424} {arXiv:1604.01424 [astro-ph.CO]}
  \BibitemShut {NoStop}%
\bibitem [{\citenamefont {Ananthaswamy}(2018)}]{Ananthaswamy:2018oh}%
  \BibitemOpen
  \bibfield  {author} {\bibinfo {author} {\bibfnamefont {A.}~\bibnamefont
  {Ananthaswamy}},\ }\href {https://doi.org/10.1073/pnas.1811473115} {\bibfield
   {journal} {\bibinfo  {journal} {Proc. Nat. Acad. Sci.}\ }\textbf {\bibinfo
  {volume} {115}},\ \bibinfo {pages} {9810} (\bibinfo {year}
  {2018})}\BibitemShut {NoStop}%
\bibitem [{\citenamefont {Di~Valentino}(2022)}]{DiValentino:2022fjm}%
  \BibitemOpen
  \bibfield  {author} {\bibinfo {author} {\bibfnamefont {E.}~\bibnamefont
  {Di~Valentino}},\ }\href {https://doi.org/10.3390/universe8080399} {\bibfield
   {journal} {\bibinfo  {journal} {Universe}\ }\textbf {\bibinfo {volume}
  {8}},\ \bibinfo {pages} {399} (\bibinfo {year} {2022})}\BibitemShut {NoStop}%
\bibitem [{\citenamefont {Hu}\ and\ \citenamefont {Wang}(2023)}]{Hu:2023jqc}%
  \BibitemOpen
  \bibfield  {author} {\bibinfo {author} {\bibfnamefont {J.}~\bibnamefont
  {Hu}}\ and\ \bibinfo {author} {\bibfnamefont {F.~Y.}\ \bibnamefont {Wang}},\
  }\href {https://doi.org/10.3390/universe9020094} {\bibfield  {journal}
  {\bibinfo  {journal} {Universe}\ }\textbf {\bibinfo {volume} {9}},\ \bibinfo
  {pages} {94} (\bibinfo {year} {2023})},\ \Eprint
  {https://arxiv.org/abs/2302.05709} {arXiv:2302.05709 [astro-ph.CO]}
  \BibitemShut {NoStop}%
\bibitem [{\citenamefont {Coley}\ and\ \citenamefont
  {Ellis}(2019)}]{coleyellis}%
  \BibitemOpen
  \bibfield  {author} {\bibinfo {author} {\bibfnamefont {A.~A.}\ \bibnamefont
  {Coley}}\ and\ \bibinfo {author} {\bibfnamefont {G.~F.~R.}\ \bibnamefont
  {Ellis}},\ }\href@noop {} {\bibfield  {journal} {\bibinfo  {journal}
  {Classical and Quantum Gravity}\ }\textbf {\bibinfo {volume} {37}},\ \bibinfo
  {pages} {013001} (\bibinfo {year} {2019})}\BibitemShut {NoStop}%
\bibitem [{\citenamefont {Coley}(2023)}]{ColeyJWST2023}%
  \BibitemOpen
  \bibfield  {author} {\bibinfo {author} {\bibfnamefont {A.~A.}\ \bibnamefont
  {Coley}},\ }\href@noop {} {\bibfield  {journal} {\bibinfo  {journal} {ArXiv
  preprint}\ } (\bibinfo {year} {2023})},\ \Eprint
  {https://arxiv.org/abs/gr-qc/2312.14738} {gr-qc/2312.14738} \BibitemShut
  {NoStop}%
\bibitem [{\citenamefont {Hohmann}\ \emph {et~al.}(2019)\citenamefont
  {Hohmann}, \citenamefont {J\"arv}, \citenamefont {Kr\v{s}\v{s}\'ak},\ and\
  \citenamefont {Pfeifer}}]{Hohmann:2019nat}%
  \BibitemOpen
  \bibfield  {author} {\bibinfo {author} {\bibfnamefont {M.}~\bibnamefont
  {Hohmann}}, \bibinfo {author} {\bibfnamefont {L.}~\bibnamefont {J\"arv}},
  \bibinfo {author} {\bibfnamefont {M.}~\bibnamefont {Kr\v{s}\v{s}\'ak}},\ and\
  \bibinfo {author} {\bibfnamefont {C.}~\bibnamefont {Pfeifer}},\ }\href
  {https://doi.org/10.1103/PhysRevD.100.084002} {\bibfield  {journal} {\bibinfo
   {journal} {Phys. Rev. D}\ }\textbf {\bibinfo {volume} {100}},\ \bibinfo
  {pages} {084002} (\bibinfo {year} {2019})},\ \Eprint
  {https://arxiv.org/abs/1901.05472} {arXiv:1901.05472 [gr-qc]} \BibitemShut
  {NoStop}%
\bibitem [{\citenamefont {Coley}\ \emph {et~al.}(2020)\citenamefont {Coley},
  \citenamefont {van~den Hoogen},\ and\ \citenamefont
  {McNutt}}]{Coley:2019zld}%
  \BibitemOpen
  \bibfield  {author} {\bibinfo {author} {\bibfnamefont {A.~A.}\ \bibnamefont
  {Coley}}, \bibinfo {author} {\bibfnamefont {R.~J.}\ \bibnamefont {van~den
  Hoogen}},\ and\ \bibinfo {author} {\bibfnamefont {D.~D.}\ \bibnamefont
  {McNutt}},\ }\href {https://doi.org/10.1063/5.0003252} {\bibfield  {journal}
  {\bibinfo  {journal} {Journal of Mathematical Physics}\ }\textbf {\bibinfo
  {volume} {61}},\ \bibinfo {pages} {072503} (\bibinfo {year} {2020})},\
  \Eprint {https://arxiv.org/abs/1911.03893} {arXiv:1911.03893 [gr-qc]}
  \BibitemShut {NoStop}%
\bibitem [{\citenamefont {Hohmann}(2021)}]{Hohmann:2021ast}%
  \BibitemOpen
  \bibfield  {author} {\bibinfo {author} {\bibfnamefont {M.}~\bibnamefont
  {Hohmann}},\ }\href {https://doi.org/10.1103/PhysRevD.104.124077} {\bibfield
  {journal} {\bibinfo  {journal} {Phys. Rev. D}\ }\textbf {\bibinfo {volume}
  {104}},\ \bibinfo {pages} {124077} (\bibinfo {year} {2021})},\ \Eprint
  {https://arxiv.org/abs/2109.01525} {arXiv:2109.01525 [gr-qc]} \BibitemShut
  {NoStop}%
\bibitem [{\citenamefont {Pfeifer}(2022)}]{Pfeifer:2022txm}%
  \BibitemOpen
  \bibfield  {author} {\bibinfo {author} {\bibfnamefont {C.}~\bibnamefont
  {Pfeifer}},\ }\href@noop {} {\bibfield  {journal} {\bibinfo  {journal} {A
  quick guide to spacetime symmetry and symmetric solutions in teleparallel
  gravity}\ } (\bibinfo {year} {2022})},\ \Eprint
  {https://arxiv.org/abs/2201.04691} {arXiv:2201.04691 [gr-qc]} \BibitemShut
  {NoStop}%
\bibitem [{\citenamefont {McNutt}\ \emph {et~al.}(2023)\citenamefont {McNutt},
  \citenamefont {Coley},\ and\ \citenamefont {{van den
  Hoogen}}}]{McNutt:2023nxmnxm}%
  \BibitemOpen
  \bibfield  {author} {\bibinfo {author} {\bibfnamefont {D.~D.}\ \bibnamefont
  {McNutt}}, \bibinfo {author} {\bibfnamefont {A.~A.}\ \bibnamefont {Coley}},\
  and\ \bibinfo {author} {\bibfnamefont {R.~J.}\ \bibnamefont {{van den
  Hoogen}}},\ }\href {https://doi.org/10.1063/5.0134596} {\bibfield  {journal}
  {\bibinfo  {journal} {J. Math. Phys.}\ }\textbf {\bibinfo {volume} {64}},\
  \bibinfo {pages} {032503} (\bibinfo {year} {2023})},\ \Eprint
  {https://arxiv.org/abs/2302.11493} {arXiv:2302.11493 [gr-qc]} \BibitemShut
  {NoStop}%
\bibitem [{\citenamefont {Obukhov}(2023)}]{obukhov2023poincare}%
  \BibitemOpen
  \bibfield  {author} {\bibinfo {author} {\bibfnamefont {Y.~N.}\ \bibnamefont
  {Obukhov}},\ }in\ \href@noop {} {\emph {\bibinfo {booktitle} {Modified and
  Quantum Gravity: From Theory to Experimental Searches on All Scales}}}\
  (\bibinfo  {publisher} {Springer},\ \bibinfo {year} {2023})\ pp.\ \bibinfo
  {pages} {105--143}\BibitemShut {NoStop}%
\bibitem [{\citenamefont {Hehl}\ \emph {et~al.}(1995)\citenamefont {Hehl},
  \citenamefont {McCrea}, \citenamefont {Mielke},\ and\ \citenamefont
  {Ne'eman}}]{Hehl_McCrea_Mielke_Neeman1995}%
  \BibitemOpen
  \bibfield  {author} {\bibinfo {author} {\bibfnamefont {F.~W.}\ \bibnamefont
  {Hehl}}, \bibinfo {author} {\bibfnamefont {J.~D.}\ \bibnamefont {McCrea}},
  \bibinfo {author} {\bibfnamefont {E.~W.}\ \bibnamefont {Mielke}},\ and\
  \bibinfo {author} {\bibfnamefont {Y.}~\bibnamefont {Ne'eman}},\ }\href
  {https://doi.org/10.1016/0370-1573(94)00111-F} {\bibfield  {journal}
  {\bibinfo  {journal} {Physics Reports}\ }\textbf {\bibinfo {volume} {258}},\
  \bibinfo {pages} {1} (\bibinfo {year} {1995})},\ \Eprint
  {https://arxiv.org/abs/gr-qc/9402012} {arXiv:gr-qc/9402012 [gr-qc]}
  \BibitemShut {NoStop}%
%%CITATION = GR-QC/9402012;%%
\bibitem [{\citenamefont {Fonseca-Neto}\ \emph {et~al.}(1996)\citenamefont
  {Fonseca-Neto}, \citenamefont {Reboucas},\ and\ \citenamefont
  {MacCallum}}]{fonseca1996algebraic}%
  \BibitemOpen
  \bibfield  {author} {\bibinfo {author} {\bibfnamefont {J.~B.}\ \bibnamefont
  {Fonseca-Neto}}, \bibinfo {author} {\bibfnamefont {M.~J.}\ \bibnamefont
  {Reboucas}},\ and\ \bibinfo {author} {\bibfnamefont {M.~A.~H.}\ \bibnamefont
  {MacCallum}},\ }\href@noop {} {\bibfield  {journal} {\bibinfo  {journal}
  {Mathematics and Computers in Simulation}\ }\textbf {\bibinfo {volume}
  {42}},\ \bibinfo {pages} {739} (\bibinfo {year} {1996})}\BibitemShut
  {NoStop}%
\bibitem [{\citenamefont {Olver}(1995)}]{olver1995equivalence}%
  \BibitemOpen
  \bibfield  {author} {\bibinfo {author} {\bibfnamefont {P.~J.}\ \bibnamefont
  {Olver}},\ }\href@noop {} {\emph {\bibinfo {title} {Equivalence, invariants
  and symmetry}}}\ (\bibinfo  {publisher} {Cambridge University Press},\
  \bibinfo {year} {1995})\BibitemShut {NoStop}%
\bibitem [{\citenamefont {{\AA}man}\ \emph {et~al.}(1998)\citenamefont
  {{\AA}man}, \citenamefont {Fonseca-Neto}, \citenamefont {MacCallum},\ and\
  \citenamefont {Rebou{\c{c}}as}}]{aaman1998riemann}%
  \BibitemOpen
  \bibfield  {author} {\bibinfo {author} {\bibfnamefont {J.~E.}\ \bibnamefont
  {{\AA}man}}, \bibinfo {author} {\bibfnamefont {J.~B.}\ \bibnamefont
  {Fonseca-Neto}}, \bibinfo {author} {\bibfnamefont {M.~A.~H.}\ \bibnamefont
  {MacCallum}},\ and\ \bibinfo {author} {\bibfnamefont {M.~J.}\ \bibnamefont
  {Rebou{\c{c}}as}},\ }\href@noop {} {\bibfield  {journal} {\bibinfo  {journal}
  {Classical and Quantum Gravity}\ }\textbf {\bibinfo {volume} {15}},\ \bibinfo
  {pages} {1089} (\bibinfo {year} {1998})}\BibitemShut {NoStop}%
\bibitem [{\citenamefont {Yano}(2020)}]{yano2020theory}%
  \BibitemOpen
  \bibfield  {author} {\bibinfo {author} {\bibfnamefont {K.}~\bibnamefont
  {Yano}},\ }\href@noop {} {\emph {\bibinfo {title} {The theory of Lie
  derivatives and its applications}}}\ (\bibinfo  {publisher} {Courier Dover
  Publications},\ \bibinfo {year} {2020})\BibitemShut {NoStop}%
\bibitem [{\citenamefont {Lichnerowicz}(1977)}]{lichnerowicz1977geometry}%
  \BibitemOpen
  \bibfield  {author} {\bibinfo {author} {\bibfnamefont {A.}~\bibnamefont
  {Lichnerowicz}},\ }\href@noop {} {\emph {\bibinfo {title} {Geometry of groups
  of transformations}}}\ (\bibinfo  {publisher} {Noordhoff International
  Publishing},\ \bibinfo {year} {1977})\BibitemShut {NoStop}%
\bibitem [{\citenamefont {Trautman}(2006)}]{Trautman:2006fp}%
  \BibitemOpen
  \bibfield  {author} {\bibinfo {author} {\bibfnamefont {A.}~\bibnamefont
  {Trautman}},\ }\href@noop {} {\bibfield  {journal} {\bibinfo  {journal}
  {Arxiv Preprint}\ } (\bibinfo {year} {2006})},\ \Eprint
  {https://arxiv.org/abs/gr-qc/0606062} {arXiv:gr-qc/0606062} \BibitemShut
  {NoStop}%
\bibitem [{\citenamefont {McNutt}\ \emph {et~al.}(2024)\citenamefont {McNutt},
  \citenamefont {Coley},\ and\ \citenamefont {van~den Hoogen}}]{McNutt2024}%
  \BibitemOpen
  \bibfield  {author} {\bibinfo {author} {\bibfnamefont {D.~D.}\ \bibnamefont
  {McNutt}}, \bibinfo {author} {\bibfnamefont {A.~A.}\ \bibnamefont {Coley}},\
  and\ \bibinfo {author} {\bibfnamefont {R.~J.}\ \bibnamefont {van~den
  Hoogen}},\ }\href@noop {} {\bibfield  {journal} {\bibinfo  {journal} {ArXiv
  preprint}\ } (\bibinfo {year} {2024})},\ \Eprint
  {https://arxiv.org/abs/gr-qc/2401.00780} {gr-qc/2401.00780} \BibitemShut
  {NoStop}%
\bibitem [{\citenamefont {Coley}\ \emph {et~al.}(2022)\citenamefont {Coley},
  \citenamefont {Van Den~Hoogen},\ and\ \citenamefont {McNutt}}]{Coley:2022}%
  \BibitemOpen
  \bibfield  {author} {\bibinfo {author} {\bibfnamefont {A.~A.}\ \bibnamefont
  {Coley}}, \bibinfo {author} {\bibfnamefont {R.~J.}\ \bibnamefont {Van
  Den~Hoogen}},\ and\ \bibinfo {author} {\bibfnamefont {D.~D.}\ \bibnamefont
  {McNutt}},\ }\href@noop {} {\bibfield  {journal} {\bibinfo  {journal}
  {Classical and Quantum Gravity}\ }\textbf {\bibinfo {volume} {39}},\ \bibinfo
  {pages} {22LT01} (\bibinfo {year} {2022})},\ \Eprint
  {https://arxiv.org/abs/2205.10719} {arXiv:2205.10719 [gr-qc]} \BibitemShut
  {NoStop}%
\bibitem [{\citenamefont {Kruglikov}\ and\ \citenamefont
  {Winther}(2016)}]{kruglikov2016}%
  \BibitemOpen
  \bibfield  {author} {\bibinfo {author} {\bibfnamefont {B.}~\bibnamefont
  {Kruglikov}}\ and\ \bibinfo {author} {\bibfnamefont {H.}~\bibnamefont
  {Winther}},\ }\href@noop {} {\bibfield  {journal} {\bibinfo  {journal} {arXiv
  preprint arXiv:1611.05334}\ } (\bibinfo {year} {2016})}\BibitemShut {NoStop}%
\end{thebibliography}%

\newpage
\section*{Appendix: Connection one-forms} 

\begin{itemize}
\item $SO(3)$:

\beq \begin{aligned}  \bomega_{12} & = W_5(t,r) \bh^1 + W_6(t,r) \bh^2, \\
\bomega_{13} & = W_7(t,r) \bh^3 + W_8(t,r) \bh^4, \\
\bomega_{14} & = - W_8(t,r) \bh^3 + W_7(t,r) \bh^3, \\
\bomega_{23} & = W_3(t,r) \bh^3 + W_4(t,r) \bh^4, \\
\bomega_{24} & = - W_4(t,r) \bh^3 + W_3(t,r) \bh^4, \\
\bomega_{34} & = W_1(t,r) \bh^1 + W_2(t,r) \bh^2 - \frac{\cot(\theta)}{\e(t,r)} \bh^4. \end{aligned} \eeq

\item $SO(1,2)$:

\beq \begin{aligned} \bomega_{12} &= W_7(t,z) \bh^2 +W_8(t,z) \bh^3, \\
\bomega_{13} & = - W_8(t,z) \bh^2 + W_7(t,z) \bh^3, \\
\bomega_{14} & = W_5(t,z) \bh^1 + W_6(t,z) \bh^4, \\
\bomega_{23} & = W_1(t,z) \bh^1 - \frac{\coth(\theta)}{\b(t,z)} \bh^3 + W_2(t,z) \bh^4, \\
\bomega_{24} & = W_3(t,z) \bh^2 + W_4(t,z) \bh^3, \\
\bomega_{34} & = -W_4(t,z) \bh^2 + W_3(t,z) \bh^3. \end{aligned} \eeq

\item $E(2)$:

\beq \begin{aligned} \bomega_{12} & = W_3(u,v) \bk^1 + W_8(u,v) \bk^2, \\
\bomega_{13} & = W_4(u,v) \bk^3 + W_7(u,v) \bk^4, \\
\bomega_{14} & = -W_7(u,v) \bk^3 + W_4(u,v) \bk^4, \\
\bomega_{23} & = W_6 (u,v) \bk^3 + W_5(u,v) \bk^4, \\
\bomega_{24} & = -W_5(u,v) \bk^3 +W_6(u,v) \bk^4, \\
\bomega_{34} & = W_1(u,v) \bk^1 + W_2(u,v) \bk^2 - \frac{1}{\e(u,v) r} \bk^4. \end{aligned} \eeq

\end{itemize}
%%-------------------------------------------------------------------------
%%----------------------   SECTION       ----------------------------------
%%-------------------------------------------------------------------------

\end{document}